\documentclass[prb,reprint,superscriptaddress,floatfix]{revtex4-2}
\usepackage{graphicx}
\usepackage{amssymb}
\usepackage{mathtools}
\usepackage{color}
\usepackage{url}
\usepackage[usenames,dvipsnames]{xcolor}
\usepackage[normalem]{ulem}
\usepackage{braket}
\usepackage{bm}
\usepackage{dsfont}

\def\da{\downarrow}
\def\ua{\uparrow}

\graphicspath{{./}}

\begin{document}

\title{Prethermalization in the PXP Model under Continuous Quasiperiodic Driving}

\author{Pinaki Dutta}
\affiliation{Department of Physics, Indian Institute of Technology Kharagpur, Kharagpur - 721 302, India}%

\author{Sayan Choudhury}
\email{sayanchoudhury@hri.res.in}
\affiliation{Harish-Chandra Research Institute, a CI of Homi Bhabha National Institute, Chhatnag Road, Jhunsi, Allahabad 211019}%

\author{Vishwanath Shukla}
\email{vishwanath.shukla@phy.iitkgp.ac.in}
\affiliation{Department of Physics, Indian Institute of Technology Kharagpur, Kharagpur - 721 302, India}%

\date{\today}
\begin{abstract} 
Motivated by recent experiments realizing long-lived non-equilibrium states in aperiodically driven quantum many-body systems, we investigate the dynamics of a quasiperiodically driven Rydberg atom chain in the strong Rydberg blockage regime. In this regime, the system is kinetically constrained and the `PXP' model describes its dynamics. Even without driving, the PXP model exhibits many-body scarring and resultant persistent oscillations for dynamics originating from the N\'{e}el-ordered initial state. We demonstrate that a rich array of dynamical behaviors emerge when the system is subjected to a continuous drive. In the high-frequency regime, the system exhibits revivals and oscillations for the N\'{e}el ordered initial state both for periodic and quasi-periodic drives. We trace the origin of this non-ergodicity to an effective PXP Hamiltonian for both of these driving protocols in this regime. Furthermore, we demonstrate that the behavior of the fidelity and the entanglement entropy is non-monotonic at low frequencies in the high-amplitude regime. This leads to several re-entrant scarring transitions both for both the N\'{e}el-ordered and the fully polarized initial state. Our results demonstrate that continuous quasi-periodic drive protocols can provide a promising route to realize prethermal phases of matter in kinetically constrained systems.

\end{abstract}

\maketitle

\section{Introduction}
The dynamics of non-equilibrium quantum many-body systems has emerged as a topic of intense interest and debate in recent years~\cite{polkovnikov2011colloquium,eisert2015quantum,nandkishore2015many,lewis2019dynamics,borgonovi2016quantum}. On the one hand, these efforts have led to the development of driving protocols for the coherent control of quantum systems and the consequent discovery of intrinsically non-equilibrium phases of matter like Floquet topological insulators~\cite{cayssol2013floquet,rudner2020band} and time crystals~\cite{sacha2017time,khemani2019brief,else2020discrete,sacha2020time,zaletel2023colloquium}. On the other hand, these studies have shed light on the process of quantum thermalization~\cite{altman2018many,mori2018thermalization,ueda2020quantum} and the development of new paradigms like prethermalization~\cite{mallayya2019prethermalization,reimann2019typicality,o2024prethermal}, deep thermalization~\cite{ippoliti2022solvable,lucas2023generalized,bhore2023deep,mark2024maximum}, and non-thermal fixed points~\cite{berges2008nonthermal,nowak2011superfluid,schmied2019non,mikheev2023universal}. It is envisaged that these non-equilibrium processes would play a significant role in the development of future quantum technologies~\cite{bloch2022new,ye2024essay,fiderer2018quantum,engelhardt2024unified,yang2022variational,rosa2020ultra,rossini2020quantum}.\\

Quantum thermalization is best understood through the lens of the Eigenstate Thermalization Hypothesis (ETH)~\cite{deutsch1991quantum,srednicki1994chaos,kim2014testing,d2016quantum,deutsch2018eigenstate}. According to the ETH, every eigenstate in a non-integrable many-body system acts as a thermal ensemble. Thus, a system initially prepared in a `typical' state would exhibit thermalization~\cite{rigol2007relaxation,rigol2008thermalization,kaufman2016quantum,mori2018thermalization,reimann2016typical}. The most remarkable exceptions to the ETH include integrable systems~\cite{cassidy2011generalized,caux2012constructing,vidmar2016generalized,langen2015experimental}, where the system relaxes to a generalized Gibbs ensemble and many-body localized systems~\cite{nandkishore2015many,alet2018many,abanin2019colloquium}, where the memory of the initial state persists forever due to the presence of an extensive number of emergent local integrals of motion. However, integrability is not necessary to violate the ETH. In an intriguing development, a set of mechanisms have been recently discovered where thermalization is evaded only for special initial states. The origin of this weak ergodicity breaking can be traced to the presence of non-thermal finite energy-density eigenstates called `quantum many-body scars' (QMBSs) in these systems~\cite{bernien2017probing,turner2018quantum,turner2018weak,serbyn2021quantum,chandran2023quantum,moudgalya2022quantum,michailidis2020slow,nandy2024quantum}. Systems hosting QMBSs are characterized by the emergence of a scarred subspace that decouples from the rest of the Hilbert space without any underlying symmetry~\cite{choi2019emergent}. Remarkably, QMBS arise in several experimentally relevant systems and they present a new paradigm in the study of quantum thermalization~\cite{ho2019periodic,zhao2020quantum,bull2022tuning,lin2019exact,mark2020eta,shibata2020onsager,mohapatra2023pronounced,mondal2020chaos,you2022quantum,schecter2019weak,lee2020exact,mcclarty2020disorder,khare2020localized,chertkov2021motif,banerjee2021quantum,udupa2023weak,biswas2022scars,desaules2023weak,mukherjee2021constraint,chen2020persistent}.  \\

One of the most well-studied studied systems that host  QMBS is the PXP model which describes the dynamics of an ultracold atom array in the strong Rydberg blockade regime~\cite{turner2018weak,lesanovsky2012interacting}. This model is inherently non-integrable and consequently most initial states exhibit thermalization. However, the system can exhibit persistent oscillations when the system is initially prepared in a N\'{e}el-ordered initial state: $\ket{{\mathbb Z}_2} = \ket{\uparrow \downarrow \uparrow \downarrow \ldots \uparrow \downarrow}$, where $\ket{\downarrow}$ and $\ket{\uparrow} $ represent the ground and Rydberg states, respectively. This ergodicity breaking and QMBS can be traced to the emergence of an approximate ${\rm SU(2)}$ algebra in this model, and the persistence of these oscillations can be enhanced by adding certain tailored perturbations which make the ${\rm SU(2)}$ algebra exact~\cite{choi2019emergent}. The effect of periodic driving on this system has recently been explored~\cite{mukherjee2020collapse,mukherjee2020restoring,mukherjee2022periodically,banerjee2024exact,deng2023using,huang2024engineering}. Remarkably, scarring can be controlled in these systems by tuning the driving frequency, leading to the evasion of an eventual ``heat death" for certain initial states; QMBS has been successfully leveraged to realize discrete time crystals for the N\'{e}el-ordered initial state~\cite{maskara2021discrete,bluvstein2021controlling,park2023subharmonic}. \\

While periodic drives provide a powerful tool to realize intrinsically non-equilibrium phases of matter, recent investigations have pointed out that a rich zoo of prethermal phases of matter (such as time quasicrystals and time rondeau crystals) can be realized by going beyond the Floquet paradigm~\cite{else2020long,he2024experimental,zhao2019floquet,zhao2023temporal,moon2024experimental,choudhury2021self,kumar2024prethermalization}.  Unfortunately, aperiodically driven systems heat up much faster than Floquet systems, thereby raising significant challenges in realizing these phases~\cite{mori2021rigorous,pilatowsky2023complete,yan2024prethermalization,cai20221}. Intriguingly, it is possible to delay this thermalization by employing structured aperiodic drives; this leads to a long-lived prethermal phase~\cite{nandy2017aperiodically,zhao2021random,zhao2022localization,zhao2022suppression}. In particular, discrete quasiperiodic drives have been harnessed to stabilize QMBS in the PXP chain~\cite{mukherjee2020restoring}. However, to the best of our knowledge, the dynamics of this system under continuous aperiodic drives have not been studied so far. In this work, we address this gap in the literature by examining the PXP model, subjected to a continuous driving protocol composed of multiple incommensurate frequencies. We determine the conditions under which the system exhibits a prethermal regime, and examine the dependence of the dynamical behavior on the initial state. Our work opens up the possibility of engineering prethermal phases of matter in aperiodically driven many-body scarred systems.\\

The rest of the paper is organized as follows. We introduce the model and the theoretical framework in sec.~\ref{modelsection}. In sec.~\ref{sec: periodic}, we explore the dynamics of the periodically driven PXP model. In sec.~\ref{sec:quasiperiodic}, we perform a detailed study of the dynamics of the PXP chain subjected to a continuous quasiperiodic drive characterized by two incommensurate frequencies. Finally, we summarize our key findings and discuss directions for future research in sec.~\ref{sec:Discussion}.

\section{Model and Theoretical Framework}
\label{modelsection}

We examine the dynamics of a one-dimensional array of ultracold atoms described by the Hamiltonian~\cite{bernien2017probing}:
\begin{equation}\label{eq:HamRydbergchain}
	H = -\Omega\sum_{i} ( \ket{g_i}\bra{r_i} + \ket{r_i}\bra{g_i}) + \lambda\sum_i n_i + \sum_{ij} V_{ij} n_i n_j,
\end{equation}
where $\Omega$ ($\lambda$) is the Rabi frequency (detuning) of the driving laser that couples the ground state  $\ket{g_i}$ and the Rydberg state $\ket{r_i}$ of an atom at site $i$, $n_i=\ket{r_i}\bra{r_i}$, and $V_{ij} = C/r_{ij}^6$ is the strength of the van der Waals interaction between atoms at site $i$ and $j$. For the experimental parameter regime considered here, $V_{i,i+1}\gg \lambda, \Omega \gg V_{i,i+2}$; in this regime, the presence of Rydberg states on two nearest neighbouring sites becomes energetically unfavourable ($\braket{n_i n_{i+1}}=0$), leading to the effective `PXP' Hamiltonian:
\begin{equation} \label{eq:PXPmodel}
	H = - \Omega \sum_i P_{i-1} \sigma^x_i P_{i+1} + \frac{\lambda}{2}\sum_i \sigma^z_i,
\end{equation}
where $\sigma^x_i = \ket{g_i}\bra{r_i} + \ket{r_i}\bra{g_i}$, $\sigma^z_i = 2 n_i -1$ and the projector $P_i=(1-\sigma^z_i)/2$ accounts for the constraint $\braket{n_i n_{i+1}}=0$. We set $\Omega=1$ for rest of the paper. Furthermore, we have set $L=24$ and employed periodic boundary conditions throughout the paper. We note that while we focus on the non-equilibrium properties of this system in this paper, the system can host a rich zoo of equilibrium phases in low-dimensional lattice geometries~\cite{semeghini2021probing,verresen2021prediction,samajdar2021quantum,samajdar2020complex,zhang2024quantum}\\

We now proceed to investigate the many-body dynamics of the Rydberg chain under continuous periodic and quasi-periodic drives. These drives are implemented by making the detuning, $\lambda$ time-dependent:
\begin{equation}\label{eq:continuousdrive}
	\lambda(t) = \lambda_0 \sum_j \sin(\omega_j t),
\end{equation}
where $\omega_j$ are the drive frequencies. To characterize the dynamics of the system, the two main quantities that we will compute are the return probability and the entanglement entropy. The return probability (or fidelity), $F(t)$ defined by:
\begin{equation} \label{eq:fidelity}
F(t)=|\braket{\psi(t)|\psi(0)}|^2
\end{equation}
provides a measure of the overlap between the wavefunction at time $t$, $\ket{\psi(t)}$ with the initial state  $\ket{\psi(0)}$; $F(t) \sim 0$ indicates thermalization. Naturally, the time-averaged fidelity is given by
\begin{equation}
	\braket{F} = \frac{1}{\tau}\int^{\tau}_0 F(t) dt,
\end{equation}
where $\tau$ is the duration of the temporal evolution under consideration. For all the calculations presented in this work, we set $\tau = 2*10^4/\Omega$ (and $\Omega = 1$ as mentioned before).\\

Finally, we examine the quantum information dynamics by analysing the half-chain entanglement entropy $S_{\rm ent}$ defined as follows:
\begin{equation}
S_{{\rm ent}} = {\rm Tr}_{R}\left[\rho_{R} \log\rho_{R}\right],
\end{equation}
where the reduced density matrix for the right half of the chain, $\rho_{R}$ is computed by tracing over the degrees of freedom of the left half of the chain
\begin{equation}
\rho_{R} (t) = {\rm Tr}_{L}[\ket{\psi(t)}\bra{\psi(t)}].
\end{equation}
The evolution of $S_{{\rm ent}}$ provides further insights into the (non-)ergodic behavior of the system and compliments the results obtained from the fidelity calculations.

\section{Continuous periodic drive}
\label{sec: periodic}

Before discussing the case of quasi-periodic driving, we discuss the dynamical evolution of the system under a continuous periodic drive in this section. These results are crucial for setting up the stage for understanding the effect of a quasi-periodic driving protocol. To this end, we consider a drive protocol with single frequency $\omega$: 
\begin{equation}\label{eq:Ham_periodic_sin_single_freq}
H = - \Omega \sum_i P_{i-1} \sigma^x_i P_{i+1} + \frac{\lambda_0}{2} \sin(\omega t)\sum_i \sigma^z_i.	
\end{equation}
To gain insights into this system, it is instructive to consider two asymptotic regimes: (a) the high-frequency regime ($\omega/\lambda_0 \sim \omega/\Omega \gg 1$) and (b) the high amplitude regime ($\lambda_0/\Omega \gg 1$); in these cases, we can approximately determine the effective Hamiltonian $H_F$ that captures stroboscopic dynamics.

\begin{figure}
	\includegraphics[width=0.97\linewidth]{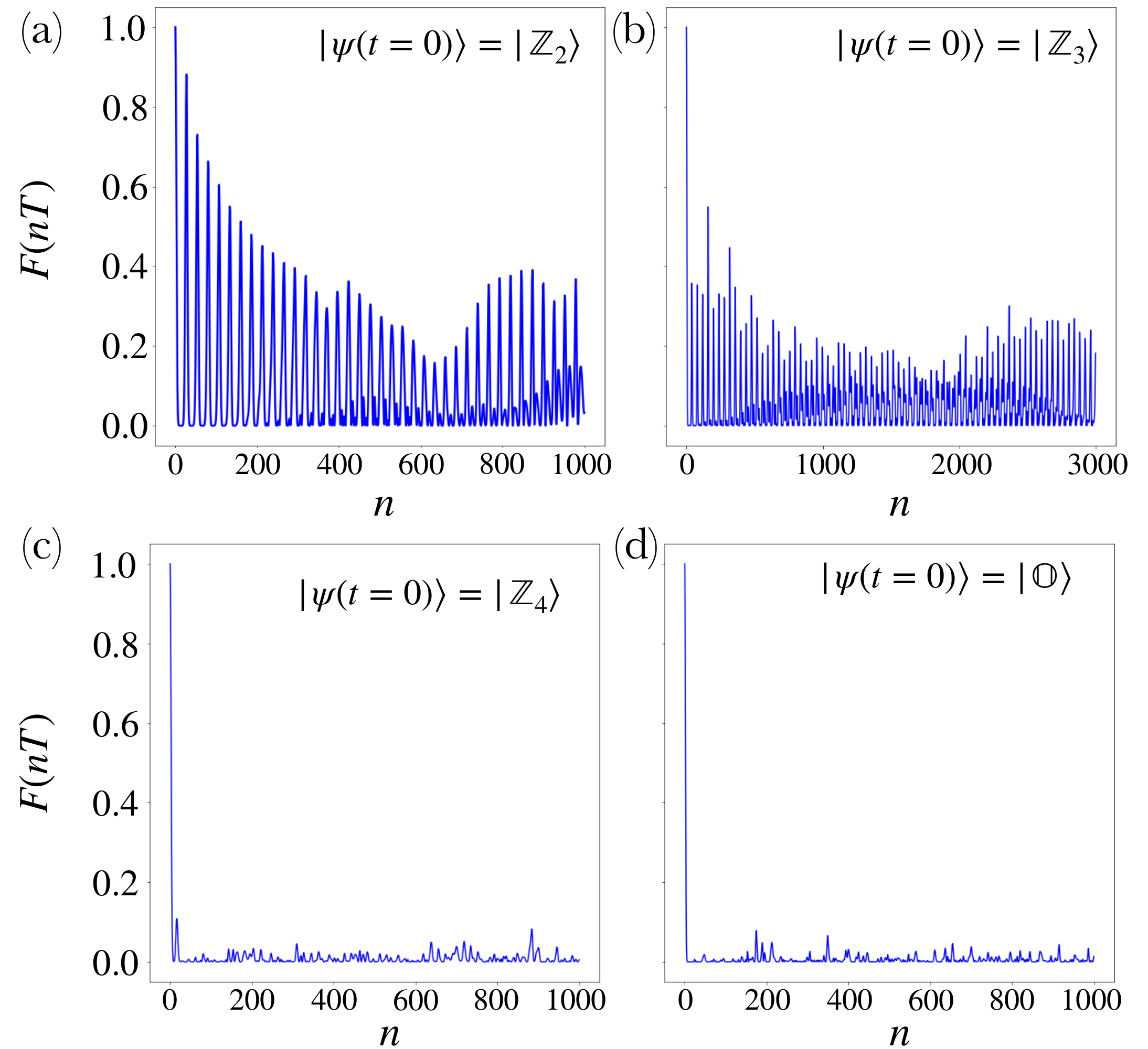}
	\caption{{\bf Dynamics of the Floquet PXP model in the high-frequency regime:} The stroboscopic fidelity $F(nT)$ for various initial states: $\ket{\mathbb{Z}_2} = \vert \uparrow \downarrow \uparrow \downarrow \ldots \uparrow \downarrow \uparrow \downarrow \rangle$, $\ket{\mathbb{Z}_3} = \vert \uparrow \downarrow \downarrow  \ldots  \uparrow \downarrow \downarrow \rangle $, $\ket{\mathbb{Z}_4} = \vert \uparrow \downarrow \downarrow \downarrow  \ldots  \uparrow \downarrow \downarrow \downarrow \rangle $, and the polarized state $\ket{\mathbb{O}} = \vert \downarrow \downarrow \ldots  \downarrow \downarrow \rangle$ when $\omega = 70$ and $\lambda_0 = 12$. In this regime, the effective Floquet Hamiltonian is of the PXP form (see eq.~\ref{floqeff}) and thus revivals and coherent oscialltions are only seen for the $\ket{\mathbb{Z}_2}$ and $\ket{\mathbb{Z}_3}$ state. All other states thermalize quickly. }
	\label{fig:FloqHighFreq}
\end{figure}

\subsection{Floquet analysis: High-Frequency Regime}

In this sub-section, we derive the Floquet Hamiltonian $H_{F}$  using high-frequency Magnus series expansion of the form:
\begin{equation}
    H_{F} = \sum_{n=0}^{\infty}T^{n}\Omega_{n},
\end{equation}
where the first term $\Omega_{0} = \frac{1}{T} \int_{0}^{T} dt_{1} H(t_{1})$ is the average Hamiltonian. Restricting to this term leads to:
\begin{equation}
\label{floqeff}
     H_{F} = - \Omega \sum_i P_{i-1} \sigma^x_i P_{i+1}
\end{equation}

Thus, at high frequencies, the N\'{e}el ordered initial state, $\ket{\mathbb{Z}_2}$ exhibits strong revivals leading to long-lived coherent oscillations, the $\ket{\mathbb{Z}_3}$ initial state ($\vert \uparrow \downarrow \downarrow  \ldots  \uparrow \downarrow \downarrow \rangle$) exhibits weaker revivals (and oscillations with a lower amplitude), and all other states thermalize very fast; the evolution of the Rydberg chain for various initial states is shown in Fig.~\ref{fig:FloqHighFreq}\\

Finally, we note that when $\lambda_0/\Omega \gg 1$, then the Floquet Hamiltonian is given by:

\begin{widetext}
\begin{equation}
\label{eq:FHamMag}
H_F = -\Omega J_0 \biggl( \frac{\lambda_0}{\omega}\biggr) \biggl[ \cos\biggl( \frac{\lambda_0}{\omega}\biggr) \sum_{i} P_{i-1} \sigma^x_i P_{i+1} - \sin\biggl( \frac{\lambda_0}{\omega}\biggr) \sum_{i} P_{i-1} \sigma^y_i P_{i+1} \biggl].
\end{equation}
\end{widetext}
Intriguingly, this form of $H_F$ persists in the low-frequency limit, as explained in the next sub-section.

\begin{figure*}
	\includegraphics[width=\linewidth]{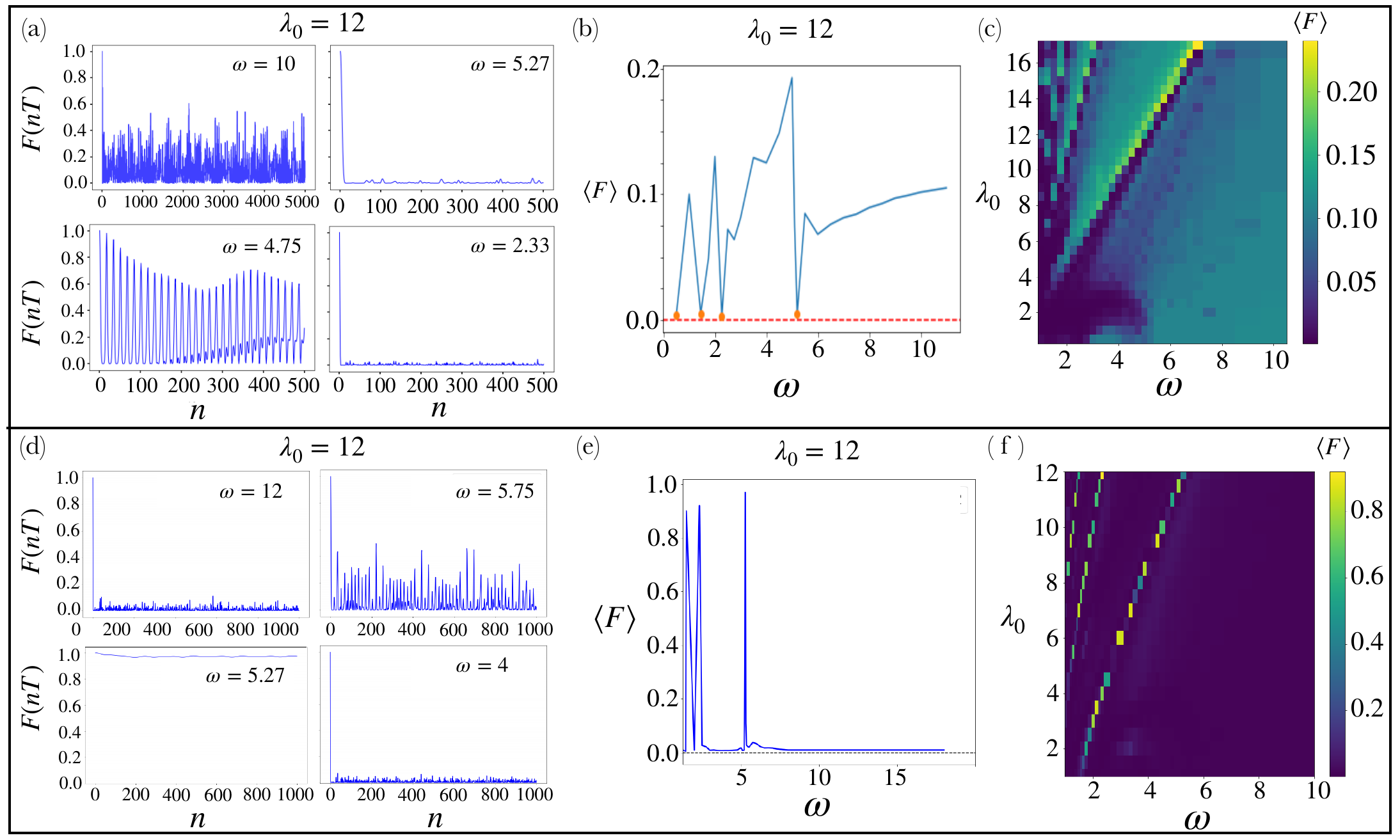}
	\caption{{\bf Dynamics of the Rydberg chain under a continuous periodic drive:} Top Panel: For the $\ket{\mathbb{Z}_2}$ state (a) Stroboscopic return probability ($F$) showing multiple transitions between non-ergodic and ergodic dynamics. (b) The time-averaged return probability $\langle F \rangle$ captures the non-monotonic response of the system as a function of $\omega$. In particular, $\langle F \rangle \sim 0$ at frequencies, $\omega$ corresponding to the zeroes of the Bessel function, $J_0 (\lambda_0/\omega)$ (see text). (c) Density plot of $\langle F \rangle$ showing the regimes with ergodic and non-ergodic dynamics in the $\lambda_0-\omega$ plane. \\
    Bottom Panel: For the $\ket{\mathbb{O}}$ state (d) Stroboscopic return probability ($F$) showing multiple transitions ergodic, oscillatory and dynamical many-body freezing regimes. (e) The time-averaged return probability,  $\langle F \rangle \sim 0$ exhibits approximate dynamical freezing appears around $\omega \sim 1.495, 2.33,$ and $5.27$ (for $\lambda_0 =12$), where the PXP terms vanish in the effective Hamiltonian and the non-PXP terms annihilate $\ket{\mathbb{O}}$ (see text). (f) The density plot of $\langle F \rangle$ shows the existence of dynamical freezing in the large $\lambda_0$ regime.}
	\label{fig:Z2_contper_sin_singlefreq}
\end{figure*}

\subsection{Floquet analysis: High Amplitude Regime}

In the high amplitude regime ($\lambda_0/\Omega \gg 1$), we employ Floquet perturbation theory to obtain the effective Floquet Hamiltonian, $H_F$ for any frequency~\cite{mukherjee2020collapse,hudomal2022driving}. We express the Hamiltonian~\eqref{eq:Ham_periodic_sin_single_freq} as
\begin{equation}
	H(t) = H_0(t) + V,
\end{equation}
where $H_0 (t) = \frac{\lambda_0}{2} \sum_i \sigma^z_{i}$ and $V=-\Omega\sum_i P_{i-1}\sigma^x_i P_{i+1}.$\\

To proceed further, we set the basis to be eigenstates of $S^z = \sum_i \sigma^z_{i}$. Hence, $H_0(t) \ket{n}= E_n(t) \ket{n}$ and $\braket{m|n}=\delta_{mn}$, where $\ket{n}$ is the set of eigenstates of $S^z$ and $E_n (t) = \bra{n} H_0(t) \ket{n}$. 

Thus, the Schr\"{o}dinger equation for $V=0:$  
\begin{equation}\label{eq:SE1}
	i\hbar \frac{\partial}{\partial t}\ket{n(t)} = H(t) \ket{n(t)},
\end{equation} 
gives the exact solution as
\begin{equation} \label{eq:SE1_exactsol}
	\ket{n(t)} = e^{-i\int^t_0 E_n(t') dt'}\ket{n(0)}.
\end{equation}
Following \cite{mukherjee2020collapse,mukherjee2022periodically}, to first order in $V$, we obtain
\begin{equation}
\label{eq:degpert_HF}
	\braket{m|H_F|n} = \frac{\braket{m|V|n}}{T}\int^T_0 dt \, e^{i \int^t_0 dt' \, [E_m(t') - E_n(t')]} .
\end{equation}
\\
We now explicitly take into account the form of the continuous periodic drive $\lambda(t)=\lambda_0 \sin(\omega t)$. Let us consider two states $\ket{m}$ and $\ket{n}$, such that $\braket{m|V|n} \neq 0$. Since, $V\ket{m} \sim \ket{m+1}+\ket{m-1}$, we find that $E_m(t)-E_n(t)=\pm \lambda(t)$. It is easy to verify that 
\begin{equation}
    \exp\left[i \int^T_0 dt \, [E_m(t) - E_n(t)]\right] = 1 \nonumber.
\end{equation} 
Hence, we must employ degenerate perturbation theory to obtain $H_F$. \\

We use the expression $\sigma^x = \tilde{\sigma}^+ + \tilde{\sigma}^-$, where $\tilde{\sigma}^{\pm}\ket{n}=\ket{n\pm 1}$, to obtain $\braket{m|V|n}=-\Omega$. Now we can use Eq.\eqref{eq:degpert_HF} to obtain
\begin{widetext}
\begin{equation}
	(H_F)_{mn} = \braket{m|H_F|n} = -\frac{\Omega}{T}\int^T_0 dt \, e^{i a \lambda_0 \int^t_0 \sin(\omega t') \, dt'} = -\Omega e^{i a \lambda_0/\omega} J_0\biggl( \frac{a \lambda_0}{\omega} \biggr),
\end{equation}
\end{widetext}
where $a=\pm 1$ and $J_0(x)$ is the Bessel function of zeroth order. Therefore, we can write the Floquet Hamiltonian, up to first order in $\Omega$, as
\begin{equation}
    H^1_F = \sum_{s=\pm 1} \sum_j \sum_m \ket{m} (H_F)_{m,m+s} \bra{m+s}, \nonumber
\end{equation}
which can be expressed as
\begin{widetext}
\begin{equation}
\label{eq:FHam}
    H^1_F = -\Omega J_0 \biggl( \frac{\lambda_0}{\omega}\biggr) \biggl[ \cos\biggl( \frac{\lambda_0}{\omega}\biggr) \sum_{j} P_{j-1} \sigma^x_i P_{j+1} - \sin\biggl( \frac{\lambda_0}{\omega}\biggr) \sum_{j} P_{j-1} \sigma^y_i P_{j+1} \biggl].
\end{equation}
\end{widetext}
We note that at large drive frequencies ($T\to 0$), $J_{0}(\lambda_{0}/\omega)\rightarrow 1$; consequently, the second term vanishes and $H^1_F$ reduces to the original PXP Hamiltonian. \\

Another interesting feature is the appearance of long-range non-PXP terms in the effective Hamiltonian, if the higher powers of $V$ is retained in the perturbation theory~\cite{mukherjee2020dynamics,hudomal2022driving}:
\begin{equation}\label{eq:Dyson}
	H_{F} = H_F^1 +   O(\Omega^3/\lambda^3) \sum[\sigma_{i-1}^{+}\sigma_{i+1}^{+}\sigma_i^{-}] + \hdots    
\end{equation}
Thus, at the zeros of $J_{0}(x)$ (i.e. $x = \lambda_{0}/\omega \approx 2.40, \,5.5, \, 8.65, \, 11.79, \,14.93, \,18.07 \ldots$), the dynamics will be governed by non-PXP terms. The competition between PXP and non-PXP like terms can lead to several reentrant transitions between the ergodic and scarred regimes. We examine these features in details by studying the time evolution of the system initialized in the N\'eel-ordered $\ket{\mathbb{Z}_2}$ and the fully polarized $\ket{\mathbb{O}}$ state.

\subsubsection{Dynamics of the N\'eel-ordered initial state}

We study the dynamics of the Rydberg chain, when it is initially prepared in the N\'eel state $\ket{\psi(0)} = \ket{\mathbb{Z}_2}=\ket{\ua \da \ua \ldots}$, under a continuous periodic drive with a single frequency for $\Omega=1$ and $\lambda=12$. In this large amplitude regime, we observe a non-monotonic behavior of the stroboscopic fidelity at low frequencies (see Fig.~\ref{fig:Z2_contper_sin_singlefreq}). In particular, the system exhibits non-monotonic reentrant transitions between completely ergodic and ergodicity-breaking dynamics as a function of $\omega$. Completely ergodic dynamics ($\braket{F} \sim 0$) arises when the non-PXP term in $H_F$ dominates; this is most evident around frequencies, $\omega \sim 5.27, \, 2.33,$ and $ 1.45$, which are in close agreement with the zero points of the Bessel function  $J_{0}(\lambda_{0}/\omega)$. For $\omega \geq 5.5$, the non-thermal behavior dominates and the $F(t)$ saturates to a finite value ($\sim 0.1$) at sufficiently high drive frequencies. \\

\begin{figure}[b]
	\includegraphics[width=\linewidth]{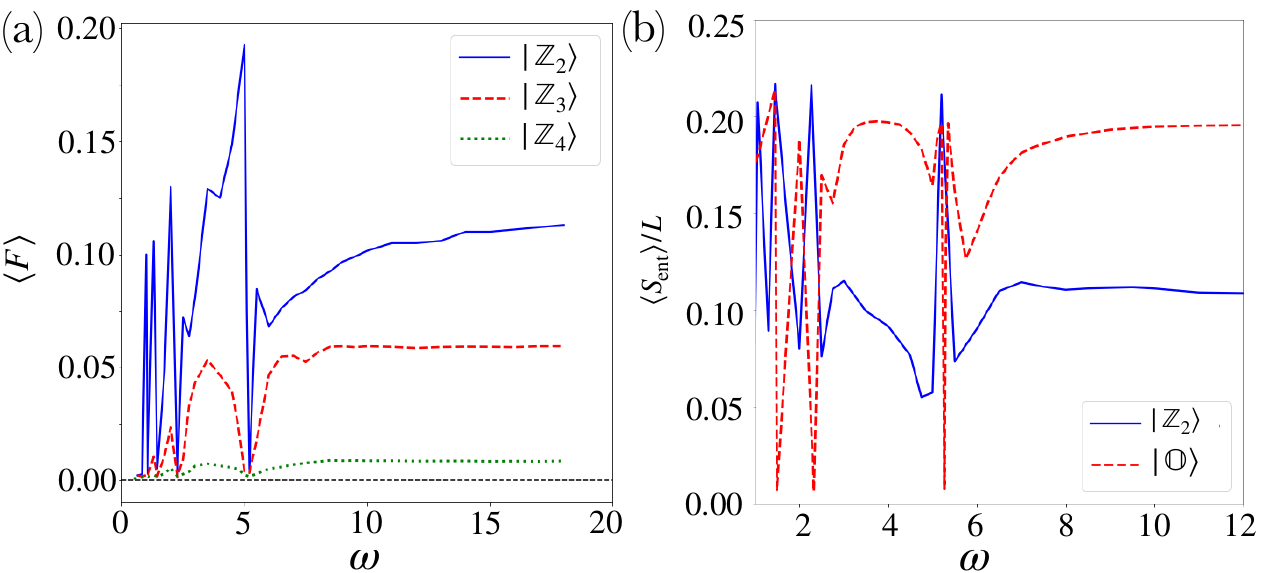}
	\caption{{\bf Comparison of the stroboscopic dynamics for various initial states:} (a) The competition between PXP and non-PXP terms in the effective Hamiltonian drives several re-entrant transitions between non-ergodic and ergodic dynamics for both the $\ket{\mathbb{Z}_2}$ and $\ket{\mathbb{Z}_3}$ states. Some signatures of this competition and the resulting transitions are also seen for the $\ket{\mathbb{Z}_4}$ state. (b) Approximate dynamical freezing leads to negligible entanglement entropy growth for the fully polarized state, $\ket{\mathbb{O}}$ when  $\omega \sim 5.27, \, 2.33,$ and $ 1.495$. Furthermore, the competition between the PXP and non-PXP terms in the effective Hamiltonian leads to complementary regimes of high and low entanglement entropy growth for the $\ket{\mathbb{Z}_2}$ and the $\ket{\mathbb{O}}$ states.}
	\label{fig:FloqEE}
\end{figure}

Interestingly, this competition between the PXP and non-PXP terms is also reflected in the dynamics of the PXP chain, when the system is initially prepared in the $\ket{\mathbb{Z}_3}$ state. Even in this situation completely ergodic dynamics ($\braket{F} \sim 0$) arises when the non-PXP term in $H_F$ dominates ($\omega \sim 5.27, \, 2.33,$ and $ 1.495$) and the system shows transitions to non-ergodic behavior between these frequencies. We note that signatures of this transition can also be seen for the $\ket{\mathbb{Z}_4}$ state, though it is considerably less prominent. Our results are shown in Fig.~\ref{fig:FloqEE}(a).  \\

\subsubsection{Dynamics of the fully polarized initial state}
We now proceed to examine the dynamics of the system when it is initially prepared in the fully polarized state $\ket{\psi(0)}=\ket{\mathbb{O}}=\ket{\da \da \da ...}$. As shown in the bottom panel of Fig.~\ref{fig:Z2_contper_sin_singlefreq}, three qualitatively different kinds of dynamical behaviors emerge as the frequency is tuned: (i) thermalization, (ii) revivals and oscillations of the fidelity, and (iii) dynamical freezing. We explain the origins of these three distinct regimes below.\\

Firstly, at high frequencies, the system thermalizes rapidly; this is expected since the dynamics in this regime is effectively described by the PXP Hamiltonian. However, when the frequency, $\omega$ is tuned to the critical values of $\omega \sim 5.27, \, 2.33,$ and $ 1.45$ (corresponding to the zeroes of $J_{0}(\lambda_{0}/\omega)$), the dynamics is completely determined by the non-PXP terms. Interestingly, these non-PXP terms annihilate the $\ket{\mathbb{O}}$; this leads to dynamical freezing, since $U(T)\rightarrow \mathds{1}$, where $\mathds{1}$ is the Identity matrix. Around these critical values of $\omega$, the PXP and non-PXP terms compete leading to coherent revivals and oscillations in the fidelity. Finally, we note that our calculations lead to an interesting conclusion: the regimes of low (high) entropy growth for the $\ket{\mathbb{O}}$ state would correspond to high (low) entropy growth for the $\ket{\mathbb{Z}_2}$ state (see Fig.~\ref{fig:FloqEE}(b)).

\section{Continuous quasi-periodic drive}
\label{sec:quasiperiodic}

In this section, we characterize the dynamics of the PXP model subjected to a quasi-periodic drive composed of two incommensurate frequencies. Analogous to the Floquet situation, we will carry out the calculations in the high-frequency and high-amplitude regimes.

\subsection{High-Frequency Regime}

We first analytically construct the effective Hamiltonian for a quasi-periodically driven Rydberg chain. We write the Hamiltonian of a system driven by $\mathcal{N}$ incommensurate frequencies as
\begin{equation}
	H = \sum_n H_n e^{i(\vec{\omega}\cdot\vec{n})t},
\end{equation}
where $\bm{\omega}=(\omega_1,\omega_2, \hdots, \omega_{\mathcal{N}})$ is the frequency vector and $\bm{n}=(n_1,n_2,\hdots,n_{\mathcal{N}})$ is a vector containing integers. Furthermore, we express the effective Hamiltonian as a perturbative series in powers of inverse drive frequency~\cite{verdeny2016quasi}, $|\omega|^{-1}$ as
\begin{equation}
	H_e(t) = \sum^{\infty}_{n=1} H^{n}_e,
\end{equation}
where $H^{(n)}_e$ is the nth order correction term. These correction terms are given by
\begin{equation}
	H^{(n)}_e = \lim_{T \to 0} \frac{1}{T}\int^{T}_0 A^{(n)}(t) \, dt,
\end{equation}
where $A^{(n)(t)}=\sum_k \frac{B_k}{k!}[ X^{(n)}_k(t) + (-1)^{k+1}Y^{(n)}_k(t) ]$ for $n\geq 2$.
$B_k$'s are the Bernoulli's numbers. The operators $X^{(n)}_k(t)$ and $Y^{(n)}_k(t)$ satisfy the following commutator recursion relations:
\begin{subequations}\label{eq:recur_quasip}
	\begin{align}
	X^{(n)}_k(t) &= \sum^{n-k}_{m=1} \biggl[ Q^{(m)}(t), X^{(n-m)}_{k-1}(t) \biggr], \\
	Y^{(n)}_k(t) &= \sum^{n-k}_{m=1} \biggl[ Q^{(m)}(t), Y^{(n-m)}_{k-1}(t) \biggr],
	\end{align}
\end{subequations}
where $Q^{(n)}(t) = \int^t_0 (A^{(n)}(t)-H^{(n)}_e)$, $X^{(1)}_0=-i H(t)$, $X^{(n)}_0=0$ for all $n\geq 2$ and $Y^{(n)}_0 = -iH^{(n)}_e$ for all $n$.\\

\begin{figure}
	\includegraphics[width=\linewidth]{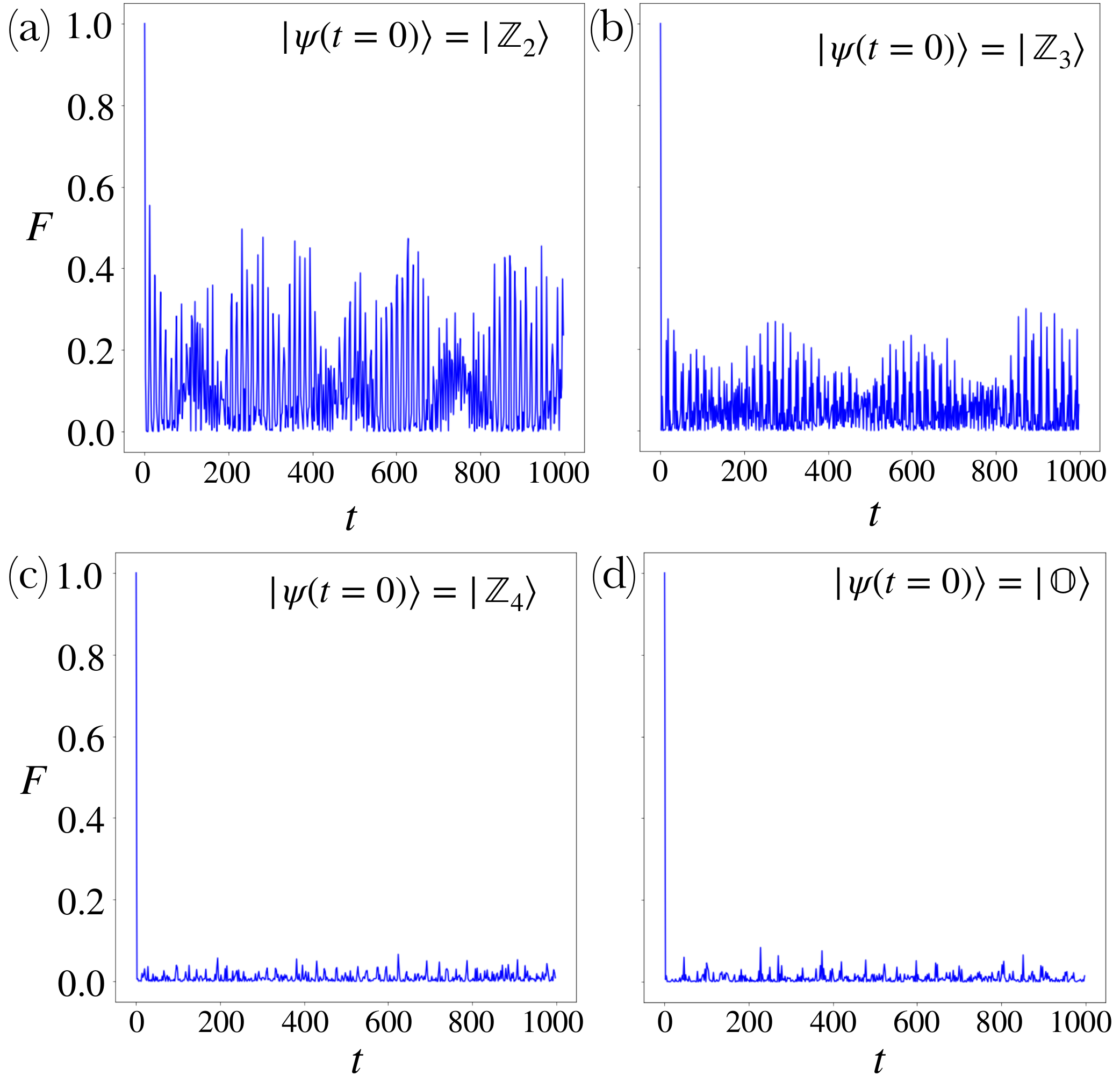}
	\caption{{\bf Dynamics of the PXP model under a quasi-periodic drive in the high-frequency regime:} The fidelity $F(t)$ for various initial states under a two-frequency drive: $\ket{\mathbb{Z}_2} = \vert \uparrow \downarrow \uparrow \downarrow \ldots \uparrow \downarrow \uparrow \downarrow \rangle$, $\ket{\mathbb{Z}_3} = \vert \uparrow \downarrow \downarrow  \ldots  \uparrow \downarrow \downarrow \rangle $, $\ket{\mathbb{Z}_4} = \vert \uparrow \downarrow \downarrow \downarrow  \ldots  \uparrow \downarrow \downarrow \downarrow \rangle $, and the polarized state $\ket{\mathbb{O}} = \vert \downarrow \downarrow \ldots  \downarrow \downarrow \rangle$. In this regime, the system is described by an effective `PXP' Hamiltonian, and thus revivals and oscillations are only seen for the $\ket{\mathbb{Z}_2}$ and $\ket{\mathbb{Z}_3}$ state. Furthermore, the revivals are much more prominent for the $\ket{\mathbb{Z}_2}$ state.}
	\label{fig:QPHighFreq}
\end{figure}

The first two terms are readily obtained as
\begin{align}
	H^{(1)}_e &= H_0, \\
	H^{(2)}_e &= \frac{1}{2} \sum_{\bm{n}\neq 0} \frac{[H_n,H_{-n}]}{\bm{\omega}\cdot \bm{n}} + \frac{[H_0,H_n]}{\bm{\omega}\cdot \bm{n}}.
\end{align}
The two-frequency drive can be expressed as 
\begin{align}
	\lambda(t) &= \lambda_0 [\sin (\omega_1 t) + \sin (\omega_2 t) ] \nonumber \\
	&= \frac{\lambda_0}{2i} \sum_{k,n=-1,1} k \exp \biggl[ i\biggl( \frac{\omega_1 - \omega_2}{2} n t  + \frac{\omega_1 + \omega_2}{2} k t\biggr) \biggr].
\end{align}
Therefore, the first and second-order correction terms are given by
\begin{align}
	H^{(1)}_e &= -\Omega \sum_i \tilde{\sigma}^x_i, \\
	\text{and} \quad	H^{(2)}_e &= \Omega \lambda_0 \biggl( \frac{1}{\omega_1} + \frac{1}{\omega_2} \biggr) \sum_i \tilde{\sigma}^y_i,
\end{align}
respectively. \\

\begin{figure*}
	\includegraphics[width=\linewidth]{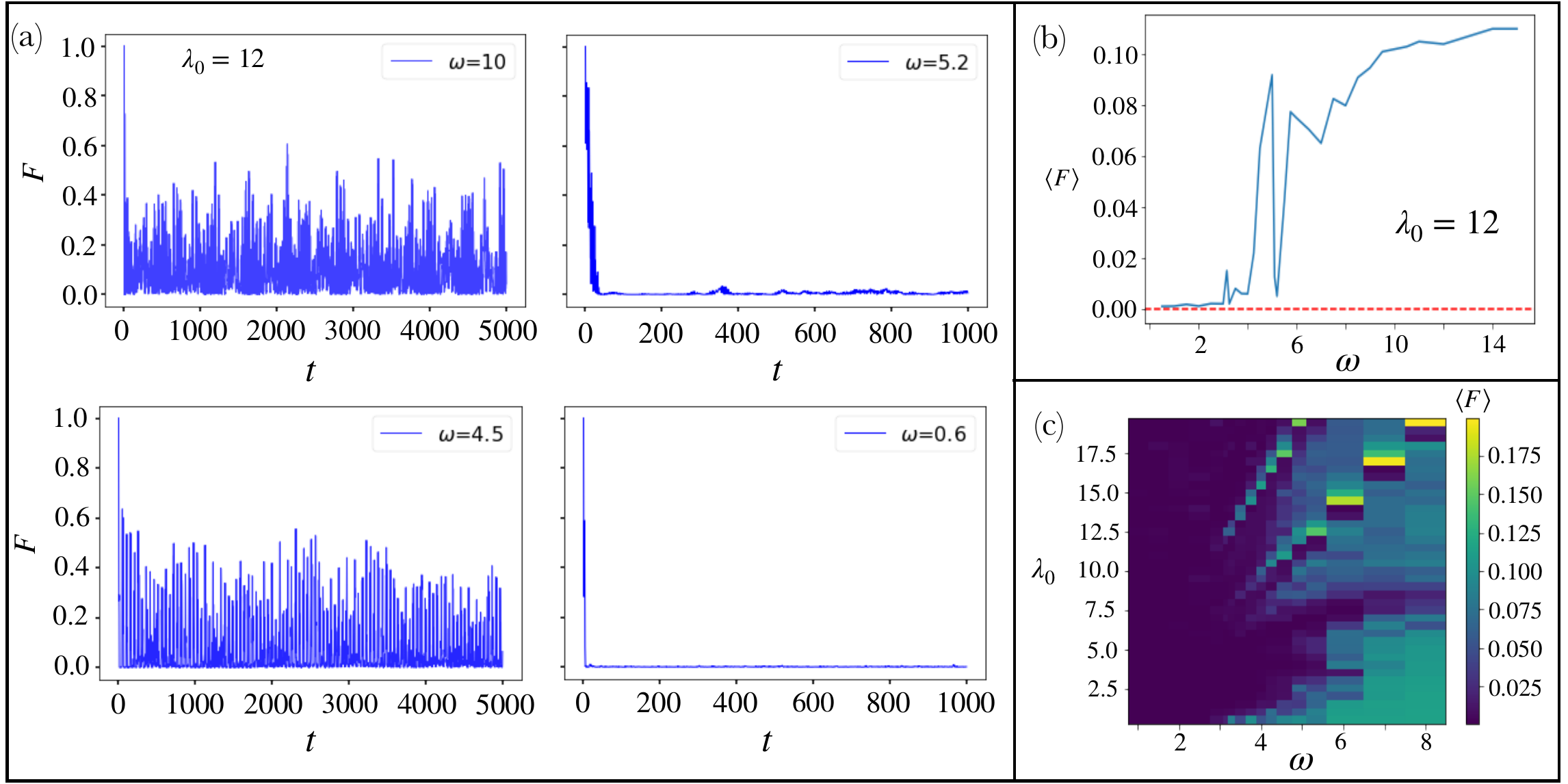}
	\caption{{\bf Dynamics of the $\ket{\mathbb{Z}_2}$ state under a continuous quasiperiodic drive:} (a) The return probability, $F$ shows a non-monotonic behavior with transitions between ergodic and non-ergodic dynamics as a function of $\omega$. (b) The time-averaged fidelity, $\langle F \rangle$ initially decreases with the fidelity up to $\omega \sim 5.2$, but then it increases again before finally decreasing. This kind of non-monotonic behavior is analogous to  (c) Density plot of the average fidelity, $\langle F \rangle$ in the $\lambda_0-\omega$ parameter plane. Multiple transitions between ergodic and non-ergodic dynamics appear at large values of $\lambda_0$. }
	\label{fig:Z2_contqper_sin_2freq}
\end{figure*}

The computation of the third-order correction term is more cumbersome; we outline this calculation below.
\begin{equation}
	H^{(3)}_e = \lim_{T \to 0} \frac{1}{T} \int^{T}_{0} A^{(3)}(t) \, dt,
 \end{equation}
where 
\begin{equation} \nonumber
	A^{(3)}(t) = \frac{1}{2} \biggl[ X^{(3)}_1(t) + Y^{(3)}_1(t) \biggr] - \frac{1}{12} \biggl[ X^{(3)}_2(t) - Y^{(3)}_2(t) \biggr].
\end{equation}
The recursion relations~\eqref{eq:recur_quasip} lead to:
\begin{subequations}\label{eq:recur_quasip_order3}
\begin{align}
	X^{(3)}_1(t) &= \biggl[Q^{(1)}(t), X^{(2)}_0(t) \biggr] + \biggl[Q^{(2)}(t), X^{(1)}_0(t) \biggr], \label{eq:recur_quasip_order3_a}\\
	Y^{(3)}_1(t) &= \biggl[Q^{(1)}(t), -i H^{(2)}_e \biggr] + \biggl[Q^{(2)}(t), -i H^{(1)}_e \biggr].
\end{align}
\end{subequations}
We note that the first commutator on the right-hand side in Eq.~\eqref{eq:recur_quasip_order3_a} is zero. We can now write $X^{1}_0(t)=-i[H_0 + \sum_m H_m \exp(i\bm{\omega}\cdot\bm{m}t)]$ and need to evaluate only the second commutator in Eq.~\eqref{eq:recur_quasip_order3_a}, where $Q^{(1)}(t)$ and $Q^{(2)}(t)$ are given by
\begin{subequations}
	\begin{align}
		Q^{(1)}(t) &= -i \sum_{\bm{n\neq 0}} \frac{H_{\bm{n}}}{\bm{n}\cdot \bm{\omega}} \biggl[e^{i (\bm{\omega}\cdot \bm{n})\, t} -1 \biggr], \\
		Q^{(2)}(t) &= \frac{1}{2} \biggl[ \sum_{\bm{n}\neq 0} \frac{[H_0,H_{\bm{n}}]}{(\bm{n}\cdot \bm{\omega})^2} [ e^{i (\bm{\omega}\cdot \bm{n}) t} -1 ]  \nonumber \\
		&+ \sum_{\bm{n},\bm{m}\neq 0, m \neq n} \frac{[H_{\bm(n)},H_{\bm(m)}]}{(\bm{n}\cdot\bm{\omega})(\bm{n}+\bm{m})\cdot\bm{\omega}} [ e^{i \bm{\omega}\cdot (\bm{n}+ \bm{m}) t} - 1] \nonumber \\
		&+ \sum_{\bm{n},\bm{m}\neq 0} \frac{[H_{\bm{n}},H_{\bm{m}}]}{(\bm{n}\cdot \bm{\omega})(\bm{m}\cdot \bm{\omega})}[e^{i \bm{\omega}\cdot \bm{m} t}-1]
		  \biggr].
	\end{align}
\end{subequations}
 \\

 \begin{figure*}[t]
	\includegraphics[width= \linewidth]{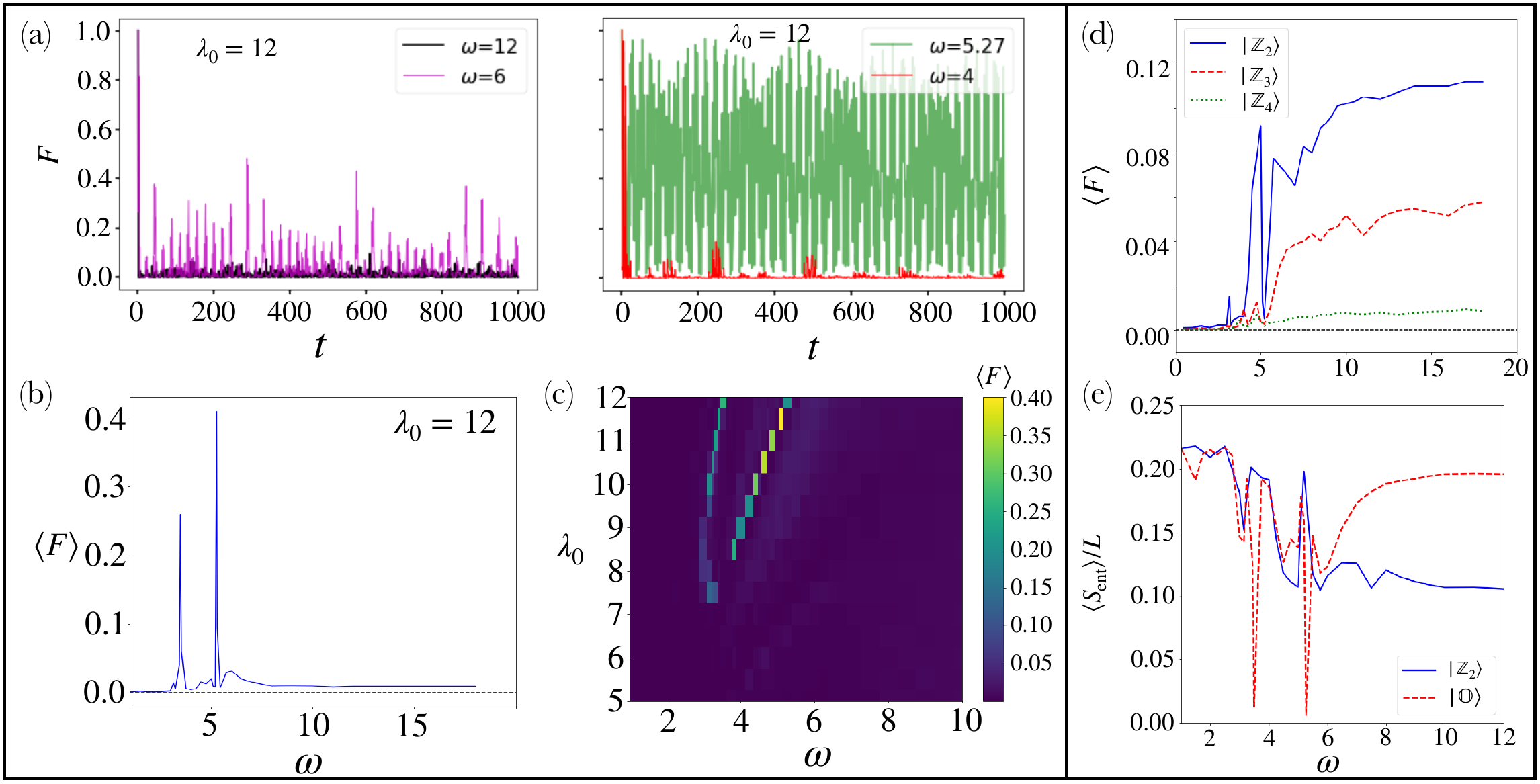}
	\caption{{\bf Dynamics of the PXP model under a quasi-periodic drive for various initial states:} Left panel: The time evolution of the fully polarized initial state. (a) The return probability $F$ shows the two different kind of dynamical regimes: (i) thermalization and (ii) revivals and coherent oscillations. (b) The time-averaged fidelity, $\langle F \rangle$ shows the non-monotonic transitions between thermalizing and ergodic regimes. In particular, revivals and coherent oscillations are seen around $\omega \sim 5.27$ and $\omega \sim 3.5$. (c) The time-averaged fidelity shows that the regimes of non-ergodic dynamics appear when $\lambda_0 > \lambda_c$, where $\lambda_c \sim 7$. \\
     Right Panel: (d) Comparison of the time-averaged fidelity, $\langle F \rangle$ of the $\ket{\mathbb{Z}_2}$, $\ket{\mathbb{Z}_3}$, and $\ket{\mathbb{Z}_4}$ states. At high frequencies, $\langle F \rangle$ for the $\ket{\mathbb{Z}_3}$ state is about half of the $\langle F \rangle$ for the $\ket{\mathbb{Z}_2}$ indicating the existence of non-ergodic dynamics at high frequencies for the  $\ket{\mathbb{Z}_3}$ state. Furthermore, $\langle F \rangle$ changes non-monotonically as a function of $\omega$ for all of these states. (e) The regimes of low (high) entanglement entropy growth for the $\ket{\mathbb{Z}_2}$ is accompanied by high (low) entanglement entropy growth of the $\ket{\mathbb{O}}$ state.}
	\label{fig:QPVacState}
\end{figure*}

For the two-frequency drive, $H_{\bm{n}}=(\lambda_0 n_2/4i)\sum_i \sigma^{z}_i$. Furthermore, in the $T \to \infty$ limit, the terms that survive are of the following form
\begin{subequations}
	\begin{align}
		A &= \frac{T}{2} \sum_{\bm{n}\neq 0}\frac{[[H_0,H_{\mathbf{n}}],H_{-\bm{n}}]}{(\bm{\omega}\cdot\bm{n})^{2}}, \\
		B &= T \sum_{\bm{n},\bm{m}\neq=0} \frac{[[H_{\bm{n}},H_0],H_{\bm{m}}]}{(\bm{\omega}\cdot\bm{n})(\bm{\omega}\cdot\bm{m})}.
	\end{align}
\end{subequations}

The substitution of $H_{\bm{n}}$ gives the following expressions for the terms A and B:
\begin{subequations}
	\begin{align}
		A &= - \frac{T\lambda^2\Omega}{4} \sum_i \tilde{\sigma}^x_i \biggl( \sum^{2}_{j=1} \frac{1}{\omega^2_j} \biggr), \\
		B &= - \frac{T\lambda^2\omega}{4} \sum_i \tilde{\sigma}^x_i \biggl(\sum^{2}_{j,k=1}\frac{4}{\omega_j\omega_k}\biggr).
	\end{align}
\end{subequations}

We obtain the third-order correction as
\begin{equation}
	H^{(3)}_e = \frac{\lambda^2_0\Omega}{8} \sum_i \tilde{\sigma}^x_i \biggl(\sum^{2}_{j,k=1}\frac{b_{j,k}}{\omega_j\omega_k}\biggr),
\end{equation}
where $b_{j,k}=5$ for $j=k$, else $b_{j,k}=4$. The form of $H^{n}_e$ ($n \in {1,2,3}$) indicates that the effective Hamiltonian is of PXP type and can qualitatively describe the scar-induced oscillations in the high-frequency limit. \\

Our analytical calculations imply that in this regime long-lived coherent oscillations will be observed for the $\ket{\mathbb{Z}_2}$ initial state; the $\ket{\mathbb{Z}_3}$ initial state would also exhibit oscillations, albeit with a smaller amplitude and other initial states thermalize rapidly. Our results are shown in Fig.~\ref{fig:QPHighFreq}. For these calculations, we have chosen a two-frequency protocol with $\omega_2=\gamma \omega$, where $\gamma=(1+\sqrt{5})/2$ is the golden ratio and $\omega$ is set to $70$. \\

\subsection{High Amplitude regime}

We next proceed to analyze the dynamics of this system in the high-amplitude regime ($\lambda_0/\omega \gg 1$). In order to compare our results to the case of periodic driving, we analyze the evolution of the system starting from the N\'eel ordered $\ket{\mathbb{Z}_2}$ state and the fully polarized state $\ket{\mathbb{O}}$.

\subsubsection{N\'eel-ordered Initial State}

We study the evolution of the system under a two-frequency quasi-periodic drive starting ($(\omega, \gamma \omega)$ from the N\'eel-ordered initial state, $\ket{\psi(t=0)} = \ket{\mathbb{Z}_2}$. In a manner analogous to the Floquet calculation, we vary $\omega$, and keep $\lambda_0=12$ fixed. Unlike the Floquet case however, we compute the fidelity and entanglement entropy at all times (and not just stroboscopically). Our results are shown in Fig.~\ref{fig:Z2_contqper_sin_2freq}; we describe the main qualitative features of the dynamics below.\\

At relatively larger frequencies ($\omega\sim 10$), the return probability, $F$ exhibits oscillations and revivals (see ~\ref{fig:Z2_contqper_sin_2freq}(a)). As the frequency is decreased further, the system initially thermalizes faster. Intriguingly, when the frequency is decreased below a threshold value ($\omega \sim 5.2)$, revivals and oscillations of the fidelity reappear; this kind of re-entrant transitions between ergodic and non-ergodic behavior is reminiscent of the Floquet situation. These transitions are efficiently captured by looking at the time-averaged fidelity, $\langle F \rangle$. On tuning to further lower frequencies ($\omega < 5.2$), $\langle F \rangle$ changes non-monotonically and the system becomes fully ergodic ($\langle F \rangle \sim 0$) when $\omega < 2$. Finally, we have examined the fate of these transitions for various values of $\lambda_0$ and $\omega$, and found that the regimes of non-ergodic dynamics become wider and more prominent with increasing $\lambda_0$ (see Fig.~\ref{fig:Z2_contqper_sin_2freq}(c)). Next, we analyze the dynamics of the system for other initial states; our results are shown in Fig.~\ref{fig:QPVacState}(a)-(e). Before proceeding further however, we note that some signatures of the non-monotonic behavior of $F$ is also seen in the case of the $\ket{\mathbb{Z}_3}$ and $\ket{\mathbb{Z}_4}$ states (see Fig.~\ref{fig:QPVacState}(d)); this is reminiscent of the dynamics of the system for a single-frequency drive.

\subsubsection{Fully Polarized Initial State}

We now examine the dynamics of the system when it is initially prepared in the fully polarized initial state, $\ket{\mathbb{O}}$. Unlike its Floquet counterpart, this system does not exhibit dynamical freezing. However, revivals and oscillations are seen around $\omega=5.27$ ($\braket{F} \sim 0.4$) and  $\omega = 3.5$ ($\braket{F} \sim 0.26$); our results are shown in Fig.~\ref{fig:QPVacState}(a)-(b). Furthermore, these oscillations disappear below a critical value of $\lambda_0 (\sim 7)$, and the system thermalizes for all frequencies (see Fig.~\ref{fig:QPVacState}(c)). This is different from the Floquet situation, where regimes of non-ergodicity persist up to much lower values of $\lambda_0$.\\

An important distinction between the Floquet dynamics of the $\ket{\mathbb{Z}_2}$ state and the $\ket{\mathbb{O}}$ state was that the regimes of high (low) entanglement growth of the $\ket{\mathbb{Z}_2}$ state correspond to low (high) entanglement growth of the $\ket{\mathbb{O}}$ state. We find that this behavior persists for the quasi-periodic driving protocol. As shown in Fig.~\ref{fig:QPVacState}(e), the entanglement growth for the $\ket{\mathbb{O}}$ state is very small ($\sim 0$) when $\omega \sim 5.27$ and $\omega \sim 3.5$. Finally, we note that for both the periodic and quasi-periodic drives, the regimes of non-ergodicity exhibit a much lower (higher) entanglement growth (average fidelity) for the $\ket{\mathbb{O}}$ state compared to the $\ket{\mathbb{Z}_2}$ state. This is quite interesting, since the $\ket{\mathbb{O}}$ state thermalizes rapidly for the static model, and the $\ket{\mathbb{Z}_2}$ state leads to revivals and oscillations. Our results highlight the remarkable properties that can be endowed on kintetically constrained systems by suitable driving protocols. 

\section{Summary and Outlook}
\label{sec:Discussion}

We have explored the non-equilibrium dynamics of a continuously driven one-dimensional Rydberg chain under two protocols: a single-frequency periodic drive and a two-frequency quasi-periodic drive. We have found that these drives considerably enrich the dynamical properties of the PXP chain. For both drive protocols the N\'eel-ordered  initial state, $\ket{\mathbb{Z}_2}$ exhibits revivals leading to oscillations in the fidelity in the high-frequency regime; this feature is also seen for the $\ket{\mathbb{Z}_3}$ state. In contrast, the system thermalizes rapidly when it is initially prepared in the fully polarized $\ket{\mathbb{O}}$ state. We provide an analytical explanation for this phenomenon by demonstrating that the PXP Hamiltonian effectively describes the dynamics of this system at high frequencies.\\

Intriguingly, in the intermediate frequencies regime the system exhibits several non-monotonic transitions between ergodic and non-ergodic dynamics, until the emergence of ergodicity in the low-frequency limit for all initial states. In the case of a periodic drive, we have traced the origin of these transitions to a competition between PXP and non-PXP terms in the effective Hamiltonian. The PXP terms dominate at higher frequencies, leading to revivals and fidelity oscillations for the  $\ket{\mathbb{Z}_2}$ state (and to a lesser extent the $\ket{\mathbb{Z}_3}$ states). However as we tune the frequency to lower values, non-PXP terms start dominating around certain frequencies. This leads to thermalization for the $\ket{\mathbb{Z}_2}$ state; however, the fully polarized $\ket{\mathbb{O}}$ state exhibits non-ergodic dynamics such as dynamical freezing (for periodic driving) and revivals (for quasi-periodic driving) at these frequencies. Thus, the regions of low entanglement growth for the $\ket{\mathbb{Z}_2}$ state is associated with high entanglement growth for the $\ket{\mathbb{O}}$ and vice-versa. We conclude that periodic and quasi-periodic drives provide a powerful tool to control non-ergodicity in the PXP model. \\

Our study serves as the starting point for several directions for future investigation. It would be interesting to explore routes to employ quasi-periodic driving to realize prethermal phases of matter that harness quantum many-body scars. Another interesting direction of research would be to study the effect of drives on non-Hermitian and dissipative versions of the PXP model. Finally, it would be worthwhile to study protocols that employ driven scarred systems for quantum-enhanced metrology.

\section*{Acknowledgements}

The authors thank Krishnendu Sengupta, Bhaskar Mukherjee, and W. Vincent Liu for discussions on the PXP model. SC thanks DST, India for support through SERB project SRG/2023/002730 and ICTS for participating in the program - Stability of Quantum Matter in and out of Equilibrium at Various Scales (code: ICTS/SQMVS2024/01). The authors acknowledge National Supercomputing Mission (NSM) for providing computing resources of ‘PARAM Shakti’ at IIT Kharagpur, which is implemented by C-DAC and supported by the Ministry of Electronics and Information Technology (MeitY) and Department of Science and Technology (DST).  VS would like to acknowledge support from the Institute Scheme for Innovative Research and Development (ISIRD), IIT Kharagpur, Grant No. IIT/SRIC/ISIRD/2021-2022/03.

\onecolumngrid
\section*{Appendix}
In the appendices, we give details about the implementations of our numerical computations (app.~\ref{sec: numerical}), the Magnus expansion for obtaining the effective Hamiltonian for the single-frequency driving (app.~\ref{sec: FloquetMagnus}).\\

\appendix
\section{Numerical methodology}
\label{sec: numerical}
The computational work was performed using PYTHON QuSpin Package~\cite{weinberg2017quspin,weinberg2019quspin} for the exact diagonalization and dynamical evolution of the initial states. We use periodic boundary conditions and take the system size $L=24$; in this case, the Hilbert space contains $103682$ states. Note that this model has a parity symmetry, as $[H,P]=0$ where $P$ is the parity operator defined through $PO_{j}P = O_{L-j-1}$. We make explicit use of both translational and parity symmetries and work in the zero momentum inversion symmetric sector, thereby reducing the Hilbert space dimension to 2359.

\section{Magnus expansion for the Floquet Hamiltonian}
\label{sec: FloquetMagnus}
In this appendix, we perform a more thorough analysis of the Floquet PXP model in the high-frequency regime. As described in the main text, Floquet Hamiltonian can be expressed in the form:

\begin{equation}
    H_{F} = \sum_{n=0}^{\infty}T^{n}\Omega_{n},
\end{equation}
where

\begin{flalign*}
\Omega_{0} & = \frac{1}{T} \int_{0}^{T} dt_{1} H(t_{1})\\ \nonumber
\Omega_{1} & = \frac{1}{2 i T^{2}} \int_{0}^{T} dt_{1} \int_{0}^{t_{1}} dt_{2} [H(t_{1}),H(t_{2})]\\ \nonumber
\Omega_{2} & = -\frac{1}{6 T^{3}} \int_{0}^{T} dt_{1} \int_{0}^{t_{1}} dt_{2} \int_{0}^{t_{2}} dt_{3} ([H(t_{1}),[H(t_{2}),H(t_{3})]] + [H(t_{3}),[H(t_{2}),H(t_{1})]]) \\ \nonumber
\Omega_{3} & = -\frac{1}{12 i T^4} \int_{0}^{T} dt_{1} \int_{0}^{t_{1}} dt_{2} \int_{0}^{t_{2}} dt_{3} \int_{0}^{t_{3}} dt_{4} 
([[[H(t_{1}),H(t_{2})],H(t_{3})],H(t_{4})] + [H(t_{1}),[[H(t_{2}),H(t_{3})],H(t_{4})]] \\ \nonumber &\quad + [H(t_{1}),[H(t_{2}),[H(t_{3}),H(t_{4})]]] + [H(t_{2}),[H(t_{3}),[H(t_{4}),H(t_{1})]]])
\end{flalign*}

The calculation of $\Omega_{0}$ and $\Omega_{1}$ is straightforward, leads to the following corrections
\begin{flalign}
H_{F}^{(0)} & = -\Omega \sum_{i} \Tilde{\sigma}^{x}_{i}\\
H_{F}^{(1)} & = \frac{\Omega \lambda_{0}}{\omega} \sum_{i} \Tilde{\sigma}^{y}_{i}
\end{flalign}
where $\Tilde{\sigma}^{\alpha}_{i} = P_{i-1}\sigma^{\alpha}_i P_{i+1}$, thereby leading to $[\sigma^{j},\Tilde{\sigma}^{k}]$ = 2 i $\epsilon_{jkl}$ $\Tilde{\sigma}^{l}$. \\

The calculation of $\Omega_{2}$ involves three commutators; however, this can be simplified and leads to the following expression:
\begin{equation}
[H(t_{1}),[H(t_{2}),H(t_{3})]] + [H(t_{3}),[H(t_{2}),H(t_{1})]] = A_0 + B_0,
\end{equation}
where

\begin{flalign*}
A_0 &= -\Omega^{2} \lambda_{0} \left(\sin(\omega t_{1}) + \sin(\omega t_{3}) - 2 \sin(\omega t_{2})\right) \sum_{i}[\Tilde{\sigma}^{x}_{i},\Tilde{\sigma}^{y}_{i}] \\ 
B_0 &= - 2i \Omega \lambda_{0}^{2} \left[2 \sin(\omega t_{1}) \sin(\omega t_{3}) - \sin(\omega t_{1}) \sin(\omega t_{2}) - \sin(\omega t_{2}) \sin(\omega t_{3})\right] \sum_{i} \Tilde{\sigma}^{x}_{i}
\end{flalign*}

After performing the necessary integrations, we get
\begin{equation}
    H_{F}^{(2)} = \frac{3 \Omega}{4}  (\frac{\lambda_{0}}{\omega})^{2} \sum_{i} \Tilde{\sigma}^{x}_{i}
\end{equation}
The derivation of $\Omega_{3}$ is a bit cumbersome as it requires successive use of the commutation relation. After some simplification, we find that 
\begin{flalign*}
-\frac{1}{12 i T^4}\int_{0}^{T} dt_{1} \int_{0}^{t_{1}} dt_{2} \int_{0}^{t_{2}} dt_{3} \int_{0}^{t_{3}} dt_{4}\\  [[[H(t_{1}),H(t_{2})],H(t_{3})],H(t_{4})] & =  A_{1} + B_{1} + C_{1}
\end{flalign*}
where 
\begin{flalign*}
    A_{1} & =   -\frac{5}{288 \pi^{3}} \Omega \lambda_{0}^{3} \sum_{i} \Tilde{\sigma}^{y}_{i}\\
    B_{1} & = \frac{6-\pi^2}{144 \pi^3}\Omega^3 \lambda_{0} \sum_{i}[[\Tilde{\sigma}^{y}_{i},\Tilde{\sigma}^{x}_{i}],\Tilde{\sigma}^{x}_{i}]\\
    C_{1} & =\frac{1}{384 \pi^2} \Omega^2 \lambda_{0}^2\sum_{i}[[\Tilde{\sigma}^{y}_{i},\Tilde{\sigma}^{x}_{i}],\sigma^{z}_{i}]
\end{flalign*}
We use the commutation relation~\cite{mukherjee2020collapse}
\begin{equation*}
[\Tilde{\sigma}^{x}_{i},\Tilde{\sigma}^{y}_{i}] = i (2 \Tilde{\sigma}^{z}_{i} + \Tilde{\sigma}^{x}_{i}\Tilde{\sigma}^{x}_{i} + \Tilde{\sigma}^{y}_{i}\Tilde{\sigma}^{y}_{i} )   
\end{equation*}
to infer $C_{1}$ = 0. \\

Similarly, the other three commutators in $\Omega_{3}$ give rise to $(A_{i},B_{i},C_{i})$ for i = 2,3,4 such that
\begin{flalign*}
    A_{2} & =   -\frac{5}{192 \pi^{3}} \Omega \lambda_{0}^{3} \sum_{i} \Tilde{\sigma}^{y}_{i}\\
    B_{2} & = \frac{1}{16}\Omega^3 \lambda_{0} \sum_{i}[[\Tilde{\sigma}^{y}_{i},\Tilde{\sigma}^{x}_{i}],\Tilde{\sigma}^{x}_{i}]\\
    A_{3} & =   -\frac{5}{288 \pi^{3}} \Omega \lambda_{0}^{3} \sum_{i} \Tilde{\sigma}^{y}_{i}\\
    B_{3} & = \frac{6-\pi^2}{144 \pi^3}\Omega^3 \lambda_{0} \sum_{i}[[\Tilde{\sigma}^{y}_{i},\Tilde{\sigma}^{x}_{i}],\Tilde{\sigma}^{x}_{i}]\\
    A_{4} & =   \frac{5}{576 \pi^{3}} \Omega \lambda_{0}^{3} \sum_{i} \Tilde{\sigma}^{y}_{i}\\
    B_{4} & = \frac{-3 + 2\pi^2}{144 \pi^3}\Omega^3 \lambda_{0} \sum_{i}[[\Tilde{\sigma}^{y}_{i},\Tilde{\sigma}^{x}_{i}],\Tilde{\sigma}^{x}_{i}]\\
\end{flalign*}
where $C_{i} = 0$. 

Finally, we note that in the large drive amplitude limit ($\lambda_{0}/\Omega >>1$), $B_{i}$ will not contribute in $H_{F}^{(3)}$ as the corresponding terms will have the order $O(\Omega^3 \lambda_{0} T^{3})$. Thus, we find
\begin{equation}
    H_{F}^{(3)} = -\frac{5 \Omega}{12}  (\frac{\lambda_{0}}{\omega})^{3} \sum_{i} \Tilde{\sigma}^{y}_{i}
\end{equation}


By the same procedure, one can obtain the higher order corrections to $H_{F}$. Note that in the limit $\lambda_{0}/\Omega >>1$, it is possible to get a power series of $\lambda_{0} /\omega$ involving only $\Tilde{\sigma}^{x}_{i}$ and $\Tilde{\sigma}^{y}_{i}$. Finally, we get

\begin{flalign}
H_{F} & = -\Omega [1-\frac{3}{4} x^2 + \frac{35}{192} x^4...]\sum_{i} \Tilde{\sigma}^{y}_{i}+ \Omega [x-\frac{5}{12} x^3 + \frac{21}{320} x^5...] \sum_{i} \Tilde{\sigma}^{x}_{i} \\ \nonumber
&= -\Omega J_0 (x)\biggl[ \cos(x) \sum_{i} \Tilde{\sigma}^{x}_{i}  - \sin(x) \sum_{i} \Tilde{\sigma}^{y}_{i} \biggl]
\end{flalign}

where $x = \lambda_{0}/\omega$. Note that the above expression of $H_{F}$ is identical to that obtained by the Floquet perturbation theory.

\twocolumngrid
\bibliographystyle{apsrev4-1}
\bibliography{ref}

\begin{thebibliography}{112}%
\makeatletter
\providecommand \@ifxundefined [1]{%
 \@ifx{#1\undefined}
}%
\providecommand \@ifnum [1]{%
 \ifnum #1\expandafter \@firstoftwo
 \else \expandafter \@secondoftwo
 \fi
}%
\providecommand \@ifx [1]{%
 \ifx #1\expandafter \@firstoftwo
 \else \expandafter \@secondoftwo
 \fi
}%
\providecommand \natexlab [1]{#1}%
\providecommand \enquote  [1]{``#1''}%
\providecommand \bibnamefont  [1]{#1}%
\providecommand \bibfnamefont [1]{#1}%
\providecommand \citenamefont [1]{#1}%
\providecommand \href@noop [0]{\@secondoftwo}%
\providecommand \href [0]{\begingroup \@sanitize@url \@href}%
\providecommand \@href[1]{\@@startlink{#1}\@@href}%
\providecommand \@@href[1]{\endgroup#1\@@endlink}%
\providecommand \@sanitize@url [0]{\catcode `\\12\catcode `\$12\catcode
  `\&12\catcode `\#12\catcode `\^12\catcode `\_12\catcode `\%12\relax}%
\providecommand \@@startlink[1]{}%
\providecommand \@@endlink[0]{}%
\providecommand \url  [0]{\begingroup\@sanitize@url \@url }%
\providecommand \@url [1]{\endgroup\@href {#1}{\urlprefix }}%
\providecommand \urlprefix  [0]{URL }%
\providecommand \Eprint [0]{\href }%
\providecommand \doibase [0]{http://dx.doi.org/}%
\providecommand \selectlanguage [0]{\@gobble}%
\providecommand \bibinfo  [0]{\@secondoftwo}%
\providecommand \bibfield  [0]{\@secondoftwo}%
\providecommand \translation [1]{[#1]}%
\providecommand \BibitemOpen [0]{}%
\providecommand \bibitemStop [0]{}%
\providecommand \bibitemNoStop [0]{.\EOS\space}%
\providecommand \EOS [0]{\spacefactor3000\relax}%
\providecommand \BibitemShut  [1]{\csname bibitem#1\endcsname}%
\let\auto@bib@innerbib\@empty
\bibitem [{\citenamefont {Polkovnikov}\ \emph {et~al.}(2011)\citenamefont
  {Polkovnikov}, \citenamefont {Sengupta}, \citenamefont {Silva},\ and\
  \citenamefont {Vengalattore}}]{polkovnikov2011colloquium}%
  \BibitemOpen
  \bibfield  {author} {\bibinfo {author} {\bibfnamefont {A.}~\bibnamefont
  {Polkovnikov}}, \bibinfo {author} {\bibfnamefont {K.}~\bibnamefont
  {Sengupta}}, \bibinfo {author} {\bibfnamefont {A.}~\bibnamefont {Silva}}, \
  and\ \bibinfo {author} {\bibfnamefont {M.}~\bibnamefont {Vengalattore}},\
  }\href@noop {} {\bibfield  {journal} {\bibinfo  {journal} {Reviews of Modern
  Physics}\ }\textbf {\bibinfo {volume} {83}},\ \bibinfo {pages} {863}
  (\bibinfo {year} {2011})}\BibitemShut {NoStop}%
\bibitem [{\citenamefont {Eisert}\ \emph {et~al.}(2015)\citenamefont {Eisert},
  \citenamefont {Friesdorf},\ and\ \citenamefont
  {Gogolin}}]{eisert2015quantum}%
  \BibitemOpen
  \bibfield  {author} {\bibinfo {author} {\bibfnamefont {J.}~\bibnamefont
  {Eisert}}, \bibinfo {author} {\bibfnamefont {M.}~\bibnamefont {Friesdorf}}, \
  and\ \bibinfo {author} {\bibfnamefont {C.}~\bibnamefont {Gogolin}},\
  }\href@noop {} {\bibfield  {journal} {\bibinfo  {journal} {Nature Physics}\
  }\textbf {\bibinfo {volume} {11}},\ \bibinfo {pages} {124} (\bibinfo {year}
  {2015})}\BibitemShut {NoStop}%
\bibitem [{\citenamefont {Nandkishore}\ and\ \citenamefont
  {Huse}(2015)}]{nandkishore2015many}%
  \BibitemOpen
  \bibfield  {author} {\bibinfo {author} {\bibfnamefont {R.}~\bibnamefont
  {Nandkishore}}\ and\ \bibinfo {author} {\bibfnamefont {D.~A.}\ \bibnamefont
  {Huse}},\ }\href@noop {} {\bibfield  {journal} {\bibinfo  {journal} {Annu.
  Rev. Condens. Matter Phys.}\ }\textbf {\bibinfo {volume} {6}},\ \bibinfo
  {pages} {15} (\bibinfo {year} {2015})}\BibitemShut {NoStop}%
\bibitem [{\citenamefont {Lewis-Swan}\ \emph {et~al.}(2019)\citenamefont
  {Lewis-Swan}, \citenamefont {Safavi-Naini}, \citenamefont {Kaufman},\ and\
  \citenamefont {Rey}}]{lewis2019dynamics}%
  \BibitemOpen
  \bibfield  {author} {\bibinfo {author} {\bibfnamefont {R.}~\bibnamefont
  {Lewis-Swan}}, \bibinfo {author} {\bibfnamefont {A.}~\bibnamefont
  {Safavi-Naini}}, \bibinfo {author} {\bibfnamefont {A.}~\bibnamefont
  {Kaufman}}, \ and\ \bibinfo {author} {\bibfnamefont {A.}~\bibnamefont
  {Rey}},\ }\href@noop {} {\bibfield  {journal} {\bibinfo  {journal} {Nature
  Reviews Physics}\ }\textbf {\bibinfo {volume} {1}},\ \bibinfo {pages} {627}
  (\bibinfo {year} {2019})}\BibitemShut {NoStop}%
\bibitem [{\citenamefont {Borgonovi}\ \emph {et~al.}(2016)\citenamefont
  {Borgonovi}, \citenamefont {Izrailev}, \citenamefont {Santos},\ and\
  \citenamefont {Zelevinsky}}]{borgonovi2016quantum}%
  \BibitemOpen
  \bibfield  {author} {\bibinfo {author} {\bibfnamefont {F.}~\bibnamefont
  {Borgonovi}}, \bibinfo {author} {\bibfnamefont {F.~M.}\ \bibnamefont
  {Izrailev}}, \bibinfo {author} {\bibfnamefont {L.~F.}\ \bibnamefont
  {Santos}}, \ and\ \bibinfo {author} {\bibfnamefont {V.~G.}\ \bibnamefont
  {Zelevinsky}},\ }\href@noop {} {\bibfield  {journal} {\bibinfo  {journal}
  {Physics Reports}\ }\textbf {\bibinfo {volume} {626}},\ \bibinfo {pages} {1}
  (\bibinfo {year} {2016})}\BibitemShut {NoStop}%
\bibitem [{\citenamefont {Cayssol}\ \emph {et~al.}(2013)\citenamefont
  {Cayssol}, \citenamefont {D{\'o}ra}, \citenamefont {Simon},\ and\
  \citenamefont {Moessner}}]{cayssol2013floquet}%
  \BibitemOpen
  \bibfield  {author} {\bibinfo {author} {\bibfnamefont {J.}~\bibnamefont
  {Cayssol}}, \bibinfo {author} {\bibfnamefont {B.}~\bibnamefont {D{\'o}ra}},
  \bibinfo {author} {\bibfnamefont {F.}~\bibnamefont {Simon}}, \ and\ \bibinfo
  {author} {\bibfnamefont {R.}~\bibnamefont {Moessner}},\ }\href@noop {}
  {\bibfield  {journal} {\bibinfo  {journal} {physica status solidi
  (RRL)--Rapid Research Letters}\ }\textbf {\bibinfo {volume} {7}},\ \bibinfo
  {pages} {101} (\bibinfo {year} {2013})}\BibitemShut {NoStop}%
\bibitem [{\citenamefont {Rudner}\ and\ \citenamefont
  {Lindner}(2020)}]{rudner2020band}%
  \BibitemOpen
  \bibfield  {author} {\bibinfo {author} {\bibfnamefont {M.~S.}\ \bibnamefont
  {Rudner}}\ and\ \bibinfo {author} {\bibfnamefont {N.~H.}\ \bibnamefont
  {Lindner}},\ }\href@noop {} {\bibfield  {journal} {\bibinfo  {journal}
  {Nature reviews physics}\ }\textbf {\bibinfo {volume} {2}},\ \bibinfo {pages}
  {229} (\bibinfo {year} {2020})}\BibitemShut {NoStop}%
\bibitem [{\citenamefont {Sacha}\ and\ \citenamefont
  {Zakrzewski}(2017)}]{sacha2017time}%
  \BibitemOpen
  \bibfield  {author} {\bibinfo {author} {\bibfnamefont {K.}~\bibnamefont
  {Sacha}}\ and\ \bibinfo {author} {\bibfnamefont {J.}~\bibnamefont
  {Zakrzewski}},\ }\href@noop {} {\bibfield  {journal} {\bibinfo  {journal}
  {Reports on Progress in Physics}\ }\textbf {\bibinfo {volume} {81}},\
  \bibinfo {pages} {016401} (\bibinfo {year} {2017})}\BibitemShut {NoStop}%
\bibitem [{\citenamefont {Khemani}\ \emph {et~al.}(2019)\citenamefont
  {Khemani}, \citenamefont {Moessner},\ and\ \citenamefont
  {Sondhi}}]{khemani2019brief}%
  \BibitemOpen
  \bibfield  {author} {\bibinfo {author} {\bibfnamefont {V.}~\bibnamefont
  {Khemani}}, \bibinfo {author} {\bibfnamefont {R.}~\bibnamefont {Moessner}}, \
  and\ \bibinfo {author} {\bibfnamefont {S.}~\bibnamefont {Sondhi}},\
  }\href@noop {} {\bibfield  {journal} {\bibinfo  {journal} {arXiv preprint
  arXiv:1910.10745}\ } (\bibinfo {year} {2019})}\BibitemShut {NoStop}%
\bibitem [{\citenamefont {Else}\ \emph
  {et~al.}(2020{\natexlab{a}})\citenamefont {Else}, \citenamefont {Monroe},
  \citenamefont {Nayak},\ and\ \citenamefont {Yao}}]{else2020discrete}%
  \BibitemOpen
  \bibfield  {author} {\bibinfo {author} {\bibfnamefont {D.~V.}\ \bibnamefont
  {Else}}, \bibinfo {author} {\bibfnamefont {C.}~\bibnamefont {Monroe}},
  \bibinfo {author} {\bibfnamefont {C.}~\bibnamefont {Nayak}}, \ and\ \bibinfo
  {author} {\bibfnamefont {N.~Y.}\ \bibnamefont {Yao}},\ }\href@noop {}
  {\bibfield  {journal} {\bibinfo  {journal} {Annual Review of Condensed Matter
  Physics}\ }\textbf {\bibinfo {volume} {11}},\ \bibinfo {pages} {467}
  (\bibinfo {year} {2020}{\natexlab{a}})}\BibitemShut {NoStop}%
\bibitem [{\citenamefont {Sacha}(2020)}]{sacha2020time}%
  \BibitemOpen
  \bibfield  {author} {\bibinfo {author} {\bibfnamefont {K.}~\bibnamefont
  {Sacha}},\ }\href@noop {} {\emph {\bibinfo {title} {Time crystals}}},\ Vol.\
  \bibinfo {volume} {114}\ (\bibinfo  {publisher} {Springer},\ \bibinfo {year}
  {2020})\BibitemShut {NoStop}%
\bibitem [{\citenamefont {Zaletel}\ \emph {et~al.}(2023)\citenamefont
  {Zaletel}, \citenamefont {Lukin}, \citenamefont {Monroe}, \citenamefont
  {Nayak}, \citenamefont {Wilczek},\ and\ \citenamefont
  {Yao}}]{zaletel2023colloquium}%
  \BibitemOpen
  \bibfield  {author} {\bibinfo {author} {\bibfnamefont {M.~P.}\ \bibnamefont
  {Zaletel}}, \bibinfo {author} {\bibfnamefont {M.}~\bibnamefont {Lukin}},
  \bibinfo {author} {\bibfnamefont {C.}~\bibnamefont {Monroe}}, \bibinfo
  {author} {\bibfnamefont {C.}~\bibnamefont {Nayak}}, \bibinfo {author}
  {\bibfnamefont {F.}~\bibnamefont {Wilczek}}, \ and\ \bibinfo {author}
  {\bibfnamefont {N.~Y.}\ \bibnamefont {Yao}},\ }\href@noop {} {\bibfield
  {journal} {\bibinfo  {journal} {Reviews of Modern Physics}\ }\textbf
  {\bibinfo {volume} {95}},\ \bibinfo {pages} {031001} (\bibinfo {year}
  {2023})}\BibitemShut {NoStop}%
\bibitem [{\citenamefont {Altman}(2018)}]{altman2018many}%
  \BibitemOpen
  \bibfield  {author} {\bibinfo {author} {\bibfnamefont {E.}~\bibnamefont
  {Altman}},\ }\href@noop {} {\bibfield  {journal} {\bibinfo  {journal} {Nature
  Physics}\ }\textbf {\bibinfo {volume} {14}},\ \bibinfo {pages} {979}
  (\bibinfo {year} {2018})}\BibitemShut {NoStop}%
\bibitem [{\citenamefont {Mori}\ \emph {et~al.}(2018)\citenamefont {Mori},
  \citenamefont {Ikeda}, \citenamefont {Kaminishi},\ and\ \citenamefont
  {Ueda}}]{mori2018thermalization}%
  \BibitemOpen
  \bibfield  {author} {\bibinfo {author} {\bibfnamefont {T.}~\bibnamefont
  {Mori}}, \bibinfo {author} {\bibfnamefont {T.~N.}\ \bibnamefont {Ikeda}},
  \bibinfo {author} {\bibfnamefont {E.}~\bibnamefont {Kaminishi}}, \ and\
  \bibinfo {author} {\bibfnamefont {M.}~\bibnamefont {Ueda}},\ }\href@noop {}
  {\bibfield  {journal} {\bibinfo  {journal} {Journal of Physics B: Atomic,
  Molecular and Optical Physics}\ }\textbf {\bibinfo {volume} {51}},\ \bibinfo
  {pages} {112001} (\bibinfo {year} {2018})}\BibitemShut {NoStop}%
\bibitem [{\citenamefont {Ueda}(2020)}]{ueda2020quantum}%
  \BibitemOpen
  \bibfield  {author} {\bibinfo {author} {\bibfnamefont {M.}~\bibnamefont
  {Ueda}},\ }\href@noop {} {\bibfield  {journal} {\bibinfo  {journal} {Nature
  Reviews Physics}\ }\textbf {\bibinfo {volume} {2}},\ \bibinfo {pages} {669}
  (\bibinfo {year} {2020})}\BibitemShut {NoStop}%
\bibitem [{\citenamefont {Mallayya}\ \emph {et~al.}(2019)\citenamefont
  {Mallayya}, \citenamefont {Rigol},\ and\ \citenamefont
  {De~Roeck}}]{mallayya2019prethermalization}%
  \BibitemOpen
  \bibfield  {author} {\bibinfo {author} {\bibfnamefont {K.}~\bibnamefont
  {Mallayya}}, \bibinfo {author} {\bibfnamefont {M.}~\bibnamefont {Rigol}}, \
  and\ \bibinfo {author} {\bibfnamefont {W.}~\bibnamefont {De~Roeck}},\
  }\href@noop {} {\bibfield  {journal} {\bibinfo  {journal} {Physical Review
  X}\ }\textbf {\bibinfo {volume} {9}},\ \bibinfo {pages} {021027} (\bibinfo
  {year} {2019})}\BibitemShut {NoStop}%
\bibitem [{\citenamefont {Reimann}\ and\ \citenamefont
  {Dabelow}(2019)}]{reimann2019typicality}%
  \BibitemOpen
  \bibfield  {author} {\bibinfo {author} {\bibfnamefont {P.}~\bibnamefont
  {Reimann}}\ and\ \bibinfo {author} {\bibfnamefont {L.}~\bibnamefont
  {Dabelow}},\ }\href@noop {} {\bibfield  {journal} {\bibinfo  {journal}
  {Physical Review Letters}\ }\textbf {\bibinfo {volume} {122}},\ \bibinfo
  {pages} {080603} (\bibinfo {year} {2019})}\BibitemShut {NoStop}%
\bibitem [{\citenamefont {O’Dea}\ \emph {et~al.}(2024)\citenamefont
  {O’Dea}, \citenamefont {Burnell}, \citenamefont {Chandran},\ and\
  \citenamefont {Khemani}}]{o2024prethermal}%
  \BibitemOpen
  \bibfield  {author} {\bibinfo {author} {\bibfnamefont {N.}~\bibnamefont
  {O’Dea}}, \bibinfo {author} {\bibfnamefont {F.}~\bibnamefont {Burnell}},
  \bibinfo {author} {\bibfnamefont {A.}~\bibnamefont {Chandran}}, \ and\
  \bibinfo {author} {\bibfnamefont {V.}~\bibnamefont {Khemani}},\ }\href@noop
  {} {\bibfield  {journal} {\bibinfo  {journal} {Physical Review Letters}\
  }\textbf {\bibinfo {volume} {132}},\ \bibinfo {pages} {100401} (\bibinfo
  {year} {2024})}\BibitemShut {NoStop}%
\bibitem [{\citenamefont {Ippoliti}\ and\ \citenamefont
  {Ho}(2022)}]{ippoliti2022solvable}%
  \BibitemOpen
  \bibfield  {author} {\bibinfo {author} {\bibfnamefont {M.}~\bibnamefont
  {Ippoliti}}\ and\ \bibinfo {author} {\bibfnamefont {W.~W.}\ \bibnamefont
  {Ho}},\ }\href@noop {} {\bibfield  {journal} {\bibinfo  {journal} {Quantum}\
  }\textbf {\bibinfo {volume} {6}},\ \bibinfo {pages} {886} (\bibinfo {year}
  {2022})}\BibitemShut {NoStop}%
\bibitem [{\citenamefont {Lucas}\ \emph {et~al.}(2023)\citenamefont {Lucas},
  \citenamefont {Piroli}, \citenamefont {De~Nardis},\ and\ \citenamefont
  {De~Luca}}]{lucas2023generalized}%
  \BibitemOpen
  \bibfield  {author} {\bibinfo {author} {\bibfnamefont {M.}~\bibnamefont
  {Lucas}}, \bibinfo {author} {\bibfnamefont {L.}~\bibnamefont {Piroli}},
  \bibinfo {author} {\bibfnamefont {J.}~\bibnamefont {De~Nardis}}, \ and\
  \bibinfo {author} {\bibfnamefont {A.}~\bibnamefont {De~Luca}},\ }\href@noop
  {} {\bibfield  {journal} {\bibinfo  {journal} {Physical Review A}\ }\textbf
  {\bibinfo {volume} {107}},\ \bibinfo {pages} {032215} (\bibinfo {year}
  {2023})}\BibitemShut {NoStop}%
\bibitem [{\citenamefont {Bhore}\ \emph {et~al.}(2023)\citenamefont {Bhore},
  \citenamefont {Desaules},\ and\ \citenamefont {Papi{\'c}}}]{bhore2023deep}%
  \BibitemOpen
  \bibfield  {author} {\bibinfo {author} {\bibfnamefont {T.}~\bibnamefont
  {Bhore}}, \bibinfo {author} {\bibfnamefont {J.-Y.}\ \bibnamefont {Desaules}},
  \ and\ \bibinfo {author} {\bibfnamefont {Z.}~\bibnamefont {Papi{\'c}}},\
  }\href@noop {} {\bibfield  {journal} {\bibinfo  {journal} {Physical Review
  B}\ }\textbf {\bibinfo {volume} {108}},\ \bibinfo {pages} {104317} (\bibinfo
  {year} {2023})}\BibitemShut {NoStop}%
\bibitem [{\citenamefont {Mark}\ \emph {et~al.}(2024)\citenamefont {Mark},
  \citenamefont {Surace}, \citenamefont {Elben}, \citenamefont {Shaw},
  \citenamefont {Choi}, \citenamefont {Refael}, \citenamefont {Endres},\ and\
  \citenamefont {Choi}}]{mark2024maximum}%
  \BibitemOpen
  \bibfield  {author} {\bibinfo {author} {\bibfnamefont {D.~K.}\ \bibnamefont
  {Mark}}, \bibinfo {author} {\bibfnamefont {F.}~\bibnamefont {Surace}},
  \bibinfo {author} {\bibfnamefont {A.}~\bibnamefont {Elben}}, \bibinfo
  {author} {\bibfnamefont {A.~L.}\ \bibnamefont {Shaw}}, \bibinfo {author}
  {\bibfnamefont {J.}~\bibnamefont {Choi}}, \bibinfo {author} {\bibfnamefont
  {G.}~\bibnamefont {Refael}}, \bibinfo {author} {\bibfnamefont
  {M.}~\bibnamefont {Endres}}, \ and\ \bibinfo {author} {\bibfnamefont
  {S.}~\bibnamefont {Choi}},\ }\href@noop {} {\bibfield  {journal} {\bibinfo
  {journal} {arXiv preprint arXiv:2403.11970}\ } (\bibinfo {year}
  {2024})}\BibitemShut {NoStop}%
\bibitem [{\citenamefont {Berges}\ \emph {et~al.}(2008)\citenamefont {Berges},
  \citenamefont {Rothkopf},\ and\ \citenamefont
  {Schmidt}}]{berges2008nonthermal}%
  \BibitemOpen
  \bibfield  {author} {\bibinfo {author} {\bibfnamefont {J.}~\bibnamefont
  {Berges}}, \bibinfo {author} {\bibfnamefont {A.}~\bibnamefont {Rothkopf}}, \
  and\ \bibinfo {author} {\bibfnamefont {J.}~\bibnamefont {Schmidt}},\
  }\href@noop {} {\bibfield  {journal} {\bibinfo  {journal} {Physical Review
  Letters}\ }\textbf {\bibinfo {volume} {101}},\ \bibinfo {pages} {041603}
  (\bibinfo {year} {2008})}\BibitemShut {NoStop}%
\bibitem [{\citenamefont {Nowak}\ \emph {et~al.}(2011)\citenamefont {Nowak},
  \citenamefont {Sexty},\ and\ \citenamefont {Gasenzer}}]{nowak2011superfluid}%
  \BibitemOpen
  \bibfield  {author} {\bibinfo {author} {\bibfnamefont {B.}~\bibnamefont
  {Nowak}}, \bibinfo {author} {\bibfnamefont {D.}~\bibnamefont {Sexty}}, \ and\
  \bibinfo {author} {\bibfnamefont {T.}~\bibnamefont {Gasenzer}},\ }\href@noop
  {} {\bibfield  {journal} {\bibinfo  {journal} {Physical Review B}\ }\textbf
  {\bibinfo {volume} {84}},\ \bibinfo {pages} {020506} (\bibinfo {year}
  {2011})}\BibitemShut {NoStop}%
\bibitem [{\citenamefont {Schmied}\ \emph {et~al.}(2019)\citenamefont
  {Schmied}, \citenamefont {Mikheev},\ and\ \citenamefont
  {Gasenzer}}]{schmied2019non}%
  \BibitemOpen
  \bibfield  {author} {\bibinfo {author} {\bibfnamefont {C.-M.}\ \bibnamefont
  {Schmied}}, \bibinfo {author} {\bibfnamefont {A.~N.}\ \bibnamefont
  {Mikheev}}, \ and\ \bibinfo {author} {\bibfnamefont {T.}~\bibnamefont
  {Gasenzer}},\ }\href@noop {} {\bibfield  {journal} {\bibinfo  {journal}
  {International Journal of Modern Physics A}\ }\textbf {\bibinfo {volume}
  {34}},\ \bibinfo {pages} {1941006} (\bibinfo {year} {2019})}\BibitemShut
  {NoStop}%
\bibitem [{\citenamefont {Mikheev}\ \emph {et~al.}(2023)\citenamefont
  {Mikheev}, \citenamefont {Siovitz},\ and\ \citenamefont
  {Gasenzer}}]{mikheev2023universal}%
  \BibitemOpen
  \bibfield  {author} {\bibinfo {author} {\bibfnamefont {A.~N.}\ \bibnamefont
  {Mikheev}}, \bibinfo {author} {\bibfnamefont {I.}~\bibnamefont {Siovitz}}, \
  and\ \bibinfo {author} {\bibfnamefont {T.}~\bibnamefont {Gasenzer}},\
  }\href@noop {} {\bibfield  {journal} {\bibinfo  {journal} {The European
  Physical Journal Special Topics}\ }\textbf {\bibinfo {volume} {232}},\
  \bibinfo {pages} {3393} (\bibinfo {year} {2023})}\BibitemShut {NoStop}%
\bibitem [{\citenamefont {Bloch}\ \emph {et~al.}(2022)\citenamefont {Bloch},
  \citenamefont {Ronen}, \citenamefont {Shaham}, \citenamefont {Katz},
  \citenamefont {Volansky},\ and\ \citenamefont {Katz}}]{bloch2022new}%
  \BibitemOpen
  \bibfield  {author} {\bibinfo {author} {\bibfnamefont {I.~M.}\ \bibnamefont
  {Bloch}}, \bibinfo {author} {\bibfnamefont {G.}~\bibnamefont {Ronen}},
  \bibinfo {author} {\bibfnamefont {R.}~\bibnamefont {Shaham}}, \bibinfo
  {author} {\bibfnamefont {O.}~\bibnamefont {Katz}}, \bibinfo {author}
  {\bibfnamefont {T.}~\bibnamefont {Volansky}}, \ and\ \bibinfo {author}
  {\bibfnamefont {O.}~\bibnamefont {Katz}},\ }\href@noop {} {\bibfield
  {journal} {\bibinfo  {journal} {Science advances}\ }\textbf {\bibinfo
  {volume} {8}},\ \bibinfo {pages} {eabl8919} (\bibinfo {year}
  {2022})}\BibitemShut {NoStop}%
\bibitem [{\citenamefont {Ye}\ and\ \citenamefont
  {Zoller}(2024)}]{ye2024essay}%
  \BibitemOpen
  \bibfield  {author} {\bibinfo {author} {\bibfnamefont {J.}~\bibnamefont
  {Ye}}\ and\ \bibinfo {author} {\bibfnamefont {P.}~\bibnamefont {Zoller}},\
  }\href@noop {} {\bibfield  {journal} {\bibinfo  {journal} {Physical Review
  Letters}\ }\textbf {\bibinfo {volume} {132}},\ \bibinfo {pages} {190001}
  (\bibinfo {year} {2024})}\BibitemShut {NoStop}%
\bibitem [{\citenamefont {Fiderer}\ and\ \citenamefont
  {Braun}(2018)}]{fiderer2018quantum}%
  \BibitemOpen
  \bibfield  {author} {\bibinfo {author} {\bibfnamefont {L.~J.}\ \bibnamefont
  {Fiderer}}\ and\ \bibinfo {author} {\bibfnamefont {D.}~\bibnamefont
  {Braun}},\ }\href@noop {} {\bibfield  {journal} {\bibinfo  {journal} {Nature
  communications}\ }\textbf {\bibinfo {volume} {9}},\ \bibinfo {pages} {1351}
  (\bibinfo {year} {2018})}\BibitemShut {NoStop}%
\bibitem [{\citenamefont {Engelhardt}\ \emph {et~al.}(2024)\citenamefont
  {Engelhardt}, \citenamefont {Choudhury},\ and\ \citenamefont
  {Liu}}]{engelhardt2024unified}%
  \BibitemOpen
  \bibfield  {author} {\bibinfo {author} {\bibfnamefont {G.}~\bibnamefont
  {Engelhardt}}, \bibinfo {author} {\bibfnamefont {S.}~\bibnamefont
  {Choudhury}}, \ and\ \bibinfo {author} {\bibfnamefont {W.~V.}\ \bibnamefont
  {Liu}},\ }\href@noop {} {\bibfield  {journal} {\bibinfo  {journal} {Physical
  Review Research}\ }\textbf {\bibinfo {volume} {6}},\ \bibinfo {pages}
  {013116} (\bibinfo {year} {2024})}\BibitemShut {NoStop}%
\bibitem [{\citenamefont {Yang}\ \emph {et~al.}(2022)\citenamefont {Yang},
  \citenamefont {Pang}, \citenamefont {Chen}, \citenamefont {Jordan},\ and\
  \citenamefont {Del~Campo}}]{yang2022variational}%
  \BibitemOpen
  \bibfield  {author} {\bibinfo {author} {\bibfnamefont {J.}~\bibnamefont
  {Yang}}, \bibinfo {author} {\bibfnamefont {S.}~\bibnamefont {Pang}}, \bibinfo
  {author} {\bibfnamefont {Z.}~\bibnamefont {Chen}}, \bibinfo {author}
  {\bibfnamefont {A.~N.}\ \bibnamefont {Jordan}}, \ and\ \bibinfo {author}
  {\bibfnamefont {A.}~\bibnamefont {Del~Campo}},\ }\href@noop {} {\bibfield
  {journal} {\bibinfo  {journal} {Physical Review Letters}\ }\textbf {\bibinfo
  {volume} {128}},\ \bibinfo {pages} {160505} (\bibinfo {year}
  {2022})}\BibitemShut {NoStop}%
\bibitem [{\citenamefont {Rosa}\ \emph {et~al.}(2020)\citenamefont {Rosa},
  \citenamefont {Rossini}, \citenamefont {Andolina}, \citenamefont {Polini},\
  and\ \citenamefont {Carrega}}]{rosa2020ultra}%
  \BibitemOpen
  \bibfield  {author} {\bibinfo {author} {\bibfnamefont {D.}~\bibnamefont
  {Rosa}}, \bibinfo {author} {\bibfnamefont {D.}~\bibnamefont {Rossini}},
  \bibinfo {author} {\bibfnamefont {G.~M.}\ \bibnamefont {Andolina}}, \bibinfo
  {author} {\bibfnamefont {M.}~\bibnamefont {Polini}}, \ and\ \bibinfo {author}
  {\bibfnamefont {M.}~\bibnamefont {Carrega}},\ }\href@noop {} {\bibfield
  {journal} {\bibinfo  {journal} {Journal of High Energy Physics}\ }\textbf
  {\bibinfo {volume} {2020}},\ \bibinfo {pages} {1} (\bibinfo {year}
  {2020})}\BibitemShut {NoStop}%
\bibitem [{\citenamefont {Rossini}\ \emph {et~al.}(2020)\citenamefont
  {Rossini}, \citenamefont {Andolina}, \citenamefont {Rosa}, \citenamefont
  {Carrega},\ and\ \citenamefont {Polini}}]{rossini2020quantum}%
  \BibitemOpen
  \bibfield  {author} {\bibinfo {author} {\bibfnamefont {D.}~\bibnamefont
  {Rossini}}, \bibinfo {author} {\bibfnamefont {G.~M.}\ \bibnamefont
  {Andolina}}, \bibinfo {author} {\bibfnamefont {D.}~\bibnamefont {Rosa}},
  \bibinfo {author} {\bibfnamefont {M.}~\bibnamefont {Carrega}}, \ and\
  \bibinfo {author} {\bibfnamefont {M.}~\bibnamefont {Polini}},\ }\href@noop {}
  {\bibfield  {journal} {\bibinfo  {journal} {Physical Review Letters}\
  }\textbf {\bibinfo {volume} {125}},\ \bibinfo {pages} {236402} (\bibinfo
  {year} {2020})}\BibitemShut {NoStop}%
\bibitem [{\citenamefont {Deutsch}(1991)}]{deutsch1991quantum}%
  \BibitemOpen
  \bibfield  {author} {\bibinfo {author} {\bibfnamefont {J.~M.}\ \bibnamefont
  {Deutsch}},\ }\href@noop {} {\bibfield  {journal} {\bibinfo  {journal}
  {Physical Review A}\ }\textbf {\bibinfo {volume} {43}},\ \bibinfo {pages}
  {2046} (\bibinfo {year} {1991})}\BibitemShut {NoStop}%
\bibitem [{\citenamefont {Srednicki}(1994)}]{srednicki1994chaos}%
  \BibitemOpen
  \bibfield  {author} {\bibinfo {author} {\bibfnamefont {M.}~\bibnamefont
  {Srednicki}},\ }\href@noop {} {\bibfield  {journal} {\bibinfo  {journal}
  {Physical Review E}\ }\textbf {\bibinfo {volume} {50}},\ \bibinfo {pages}
  {888} (\bibinfo {year} {1994})}\BibitemShut {NoStop}%
\bibitem [{\citenamefont {Kim}\ \emph {et~al.}(2014)\citenamefont {Kim},
  \citenamefont {Ikeda},\ and\ \citenamefont {Huse}}]{kim2014testing}%
  \BibitemOpen
  \bibfield  {author} {\bibinfo {author} {\bibfnamefont {H.}~\bibnamefont
  {Kim}}, \bibinfo {author} {\bibfnamefont {T.~N.}\ \bibnamefont {Ikeda}}, \
  and\ \bibinfo {author} {\bibfnamefont {D.~A.}\ \bibnamefont {Huse}},\
  }\href@noop {} {\bibfield  {journal} {\bibinfo  {journal} {Physical Review
  E}\ }\textbf {\bibinfo {volume} {90}},\ \bibinfo {pages} {052105} (\bibinfo
  {year} {2014})}\BibitemShut {NoStop}%
\bibitem [{\citenamefont {D'Alessio}\ \emph {et~al.}(2016)\citenamefont
  {D'Alessio}, \citenamefont {Kafri}, \citenamefont {Polkovnikov},\ and\
  \citenamefont {Rigol}}]{d2016quantum}%
  \BibitemOpen
  \bibfield  {author} {\bibinfo {author} {\bibfnamefont {L.}~\bibnamefont
  {D'Alessio}}, \bibinfo {author} {\bibfnamefont {Y.}~\bibnamefont {Kafri}},
  \bibinfo {author} {\bibfnamefont {A.}~\bibnamefont {Polkovnikov}}, \ and\
  \bibinfo {author} {\bibfnamefont {M.}~\bibnamefont {Rigol}},\ }\href@noop {}
  {\bibfield  {journal} {\bibinfo  {journal} {Advances in Physics}\ }\textbf
  {\bibinfo {volume} {65}},\ \bibinfo {pages} {239} (\bibinfo {year}
  {2016})}\BibitemShut {NoStop}%
\bibitem [{\citenamefont {Deutsch}(2018)}]{deutsch2018eigenstate}%
  \BibitemOpen
  \bibfield  {author} {\bibinfo {author} {\bibfnamefont {J.~M.}\ \bibnamefont
  {Deutsch}},\ }\href@noop {} {\bibfield  {journal} {\bibinfo  {journal}
  {Reports on Progress in Physics}\ }\textbf {\bibinfo {volume} {81}},\
  \bibinfo {pages} {082001} (\bibinfo {year} {2018})}\BibitemShut {NoStop}%
\bibitem [{\citenamefont {Rigol}\ \emph {et~al.}(2007)\citenamefont {Rigol},
  \citenamefont {Dunjko}, \citenamefont {Yurovsky},\ and\ \citenamefont
  {Olshanii}}]{rigol2007relaxation}%
  \BibitemOpen
  \bibfield  {author} {\bibinfo {author} {\bibfnamefont {M.}~\bibnamefont
  {Rigol}}, \bibinfo {author} {\bibfnamefont {V.}~\bibnamefont {Dunjko}},
  \bibinfo {author} {\bibfnamefont {V.}~\bibnamefont {Yurovsky}}, \ and\
  \bibinfo {author} {\bibfnamefont {M.}~\bibnamefont {Olshanii}},\ }\href@noop
  {} {\bibfield  {journal} {\bibinfo  {journal} {Physical Review Letters}\
  }\textbf {\bibinfo {volume} {98}},\ \bibinfo {pages} {050405} (\bibinfo
  {year} {2007})}\BibitemShut {NoStop}%
\bibitem [{\citenamefont {Rigol}\ \emph {et~al.}(2008)\citenamefont {Rigol},
  \citenamefont {Dunjko},\ and\ \citenamefont
  {Olshanii}}]{rigol2008thermalization}%
  \BibitemOpen
  \bibfield  {author} {\bibinfo {author} {\bibfnamefont {M.}~\bibnamefont
  {Rigol}}, \bibinfo {author} {\bibfnamefont {V.}~\bibnamefont {Dunjko}}, \
  and\ \bibinfo {author} {\bibfnamefont {M.}~\bibnamefont {Olshanii}},\
  }\href@noop {} {\bibfield  {journal} {\bibinfo  {journal} {Nature}\ }\textbf
  {\bibinfo {volume} {452}},\ \bibinfo {pages} {854} (\bibinfo {year}
  {2008})}\BibitemShut {NoStop}%
\bibitem [{\citenamefont {Kaufman}\ \emph {et~al.}(2016)\citenamefont
  {Kaufman}, \citenamefont {Tai}, \citenamefont {Lukin}, \citenamefont
  {Rispoli}, \citenamefont {Schittko}, \citenamefont {Preiss},\ and\
  \citenamefont {Greiner}}]{kaufman2016quantum}%
  \BibitemOpen
  \bibfield  {author} {\bibinfo {author} {\bibfnamefont {A.~M.}\ \bibnamefont
  {Kaufman}}, \bibinfo {author} {\bibfnamefont {M.~E.}\ \bibnamefont {Tai}},
  \bibinfo {author} {\bibfnamefont {A.}~\bibnamefont {Lukin}}, \bibinfo
  {author} {\bibfnamefont {M.}~\bibnamefont {Rispoli}}, \bibinfo {author}
  {\bibfnamefont {R.}~\bibnamefont {Schittko}}, \bibinfo {author}
  {\bibfnamefont {P.~M.}\ \bibnamefont {Preiss}}, \ and\ \bibinfo {author}
  {\bibfnamefont {M.}~\bibnamefont {Greiner}},\ }\href@noop {} {\bibfield
  {journal} {\bibinfo  {journal} {Science}\ }\textbf {\bibinfo {volume}
  {353}},\ \bibinfo {pages} {794} (\bibinfo {year} {2016})}\BibitemShut
  {NoStop}%
\bibitem [{\citenamefont {Reimann}(2016)}]{reimann2016typical}%
  \BibitemOpen
  \bibfield  {author} {\bibinfo {author} {\bibfnamefont {P.}~\bibnamefont
  {Reimann}},\ }\href@noop {} {\bibfield  {journal} {\bibinfo  {journal}
  {Nature communications}\ }\textbf {\bibinfo {volume} {7}},\ \bibinfo {pages}
  {10821} (\bibinfo {year} {2016})}\BibitemShut {NoStop}%
\bibitem [{\citenamefont {Cassidy}\ \emph {et~al.}(2011)\citenamefont
  {Cassidy}, \citenamefont {Clark},\ and\ \citenamefont
  {Rigol}}]{cassidy2011generalized}%
  \BibitemOpen
  \bibfield  {author} {\bibinfo {author} {\bibfnamefont {A.~C.}\ \bibnamefont
  {Cassidy}}, \bibinfo {author} {\bibfnamefont {C.~W.}\ \bibnamefont {Clark}},
  \ and\ \bibinfo {author} {\bibfnamefont {M.}~\bibnamefont {Rigol}},\
  }\href@noop {} {\bibfield  {journal} {\bibinfo  {journal} {Physical review
  letters}\ }\textbf {\bibinfo {volume} {106}},\ \bibinfo {pages} {140405}
  (\bibinfo {year} {2011})}\BibitemShut {NoStop}%
\bibitem [{\citenamefont {Caux}\ and\ \citenamefont
  {Konik}(2012)}]{caux2012constructing}%
  \BibitemOpen
  \bibfield  {author} {\bibinfo {author} {\bibfnamefont {J.-S.}\ \bibnamefont
  {Caux}}\ and\ \bibinfo {author} {\bibfnamefont {R.~M.}\ \bibnamefont
  {Konik}},\ }\href@noop {} {\bibfield  {journal} {\bibinfo  {journal}
  {Physical Review Letters}\ }\textbf {\bibinfo {volume} {109}},\ \bibinfo
  {pages} {175301} (\bibinfo {year} {2012})}\BibitemShut {NoStop}%
\bibitem [{\citenamefont {Vidmar}\ and\ \citenamefont
  {Rigol}(2016)}]{vidmar2016generalized}%
  \BibitemOpen
  \bibfield  {author} {\bibinfo {author} {\bibfnamefont {L.}~\bibnamefont
  {Vidmar}}\ and\ \bibinfo {author} {\bibfnamefont {M.}~\bibnamefont {Rigol}},\
  }\href@noop {} {\bibfield  {journal} {\bibinfo  {journal} {Journal of
  Statistical Mechanics: Theory and Experiment}\ }\textbf {\bibinfo {volume}
  {2016}},\ \bibinfo {pages} {064007} (\bibinfo {year} {2016})}\BibitemShut
  {NoStop}%
\bibitem [{\citenamefont {Langen}\ \emph {et~al.}(2015)\citenamefont {Langen},
  \citenamefont {Erne}, \citenamefont {Geiger}, \citenamefont {Rauer},
  \citenamefont {Schweigler}, \citenamefont {Kuhnert}, \citenamefont
  {Rohringer}, \citenamefont {Mazets}, \citenamefont {Gasenzer},\ and\
  \citenamefont {Schmiedmayer}}]{langen2015experimental}%
  \BibitemOpen
  \bibfield  {author} {\bibinfo {author} {\bibfnamefont {T.}~\bibnamefont
  {Langen}}, \bibinfo {author} {\bibfnamefont {S.}~\bibnamefont {Erne}},
  \bibinfo {author} {\bibfnamefont {R.}~\bibnamefont {Geiger}}, \bibinfo
  {author} {\bibfnamefont {B.}~\bibnamefont {Rauer}}, \bibinfo {author}
  {\bibfnamefont {T.}~\bibnamefont {Schweigler}}, \bibinfo {author}
  {\bibfnamefont {M.}~\bibnamefont {Kuhnert}}, \bibinfo {author} {\bibfnamefont
  {W.}~\bibnamefont {Rohringer}}, \bibinfo {author} {\bibfnamefont {I.~E.}\
  \bibnamefont {Mazets}}, \bibinfo {author} {\bibfnamefont {T.}~\bibnamefont
  {Gasenzer}}, \ and\ \bibinfo {author} {\bibfnamefont {J.}~\bibnamefont
  {Schmiedmayer}},\ }\href@noop {} {\bibfield  {journal} {\bibinfo  {journal}
  {Science}\ }\textbf {\bibinfo {volume} {348}},\ \bibinfo {pages} {207}
  (\bibinfo {year} {2015})}\BibitemShut {NoStop}%
\bibitem [{\citenamefont {Alet}\ and\ \citenamefont
  {Laflorencie}(2018)}]{alet2018many}%
  \BibitemOpen
  \bibfield  {author} {\bibinfo {author} {\bibfnamefont {F.}~\bibnamefont
  {Alet}}\ and\ \bibinfo {author} {\bibfnamefont {N.}~\bibnamefont
  {Laflorencie}},\ }\href@noop {} {\bibfield  {journal} {\bibinfo  {journal}
  {Comptes Rendus Physique}\ }\textbf {\bibinfo {volume} {19}},\ \bibinfo
  {pages} {498} (\bibinfo {year} {2018})}\BibitemShut {NoStop}%
\bibitem [{\citenamefont {Abanin}\ \emph {et~al.}(2019)\citenamefont {Abanin},
  \citenamefont {Altman}, \citenamefont {Bloch},\ and\ \citenamefont
  {Serbyn}}]{abanin2019colloquium}%
  \BibitemOpen
  \bibfield  {author} {\bibinfo {author} {\bibfnamefont {D.~A.}\ \bibnamefont
  {Abanin}}, \bibinfo {author} {\bibfnamefont {E.}~\bibnamefont {Altman}},
  \bibinfo {author} {\bibfnamefont {I.}~\bibnamefont {Bloch}}, \ and\ \bibinfo
  {author} {\bibfnamefont {M.}~\bibnamefont {Serbyn}},\ }\href@noop {}
  {\bibfield  {journal} {\bibinfo  {journal} {Reviews of Modern Physics}\
  }\textbf {\bibinfo {volume} {91}},\ \bibinfo {pages} {021001} (\bibinfo
  {year} {2019})}\BibitemShut {NoStop}%
\bibitem [{\citenamefont {Bernien}\ \emph {et~al.}(2017)\citenamefont
  {Bernien}, \citenamefont {Schwartz}, \citenamefont {Keesling}, \citenamefont
  {Levine}, \citenamefont {Omran}, \citenamefont {Pichler}, \citenamefont
  {Choi}, \citenamefont {Zibrov}, \citenamefont {Endres}, \citenamefont
  {Greiner}, \citenamefont {Vuleti\'c},\ and\ \citenamefont
  {Lukin}}]{bernien2017probing}%
  \BibitemOpen
  \bibfield  {author} {\bibinfo {author} {\bibfnamefont {H.}~\bibnamefont
  {Bernien}}, \bibinfo {author} {\bibfnamefont {S.}~\bibnamefont {Schwartz}},
  \bibinfo {author} {\bibfnamefont {A.}~\bibnamefont {Keesling}}, \bibinfo
  {author} {\bibfnamefont {H.}~\bibnamefont {Levine}}, \bibinfo {author}
  {\bibfnamefont {A.}~\bibnamefont {Omran}}, \bibinfo {author} {\bibfnamefont
  {H.}~\bibnamefont {Pichler}}, \bibinfo {author} {\bibfnamefont
  {S.}~\bibnamefont {Choi}}, \bibinfo {author} {\bibfnamefont {A.~S.}\
  \bibnamefont {Zibrov}}, \bibinfo {author} {\bibfnamefont {M.}~\bibnamefont
  {Endres}}, \bibinfo {author} {\bibfnamefont {M.}~\bibnamefont {Greiner}},
  \bibinfo {author} {\bibfnamefont {V.}~\bibnamefont {Vuleti\'c}}, \ and\
  \bibinfo {author} {\bibfnamefont {M.~D.}\ \bibnamefont {Lukin}},\ }\href@noop
  {} {\bibfield  {journal} {\bibinfo  {journal} {Nature}\ }\textbf {\bibinfo
  {volume} {551}},\ \bibinfo {pages} {579} (\bibinfo {year}
  {2017})}\BibitemShut {NoStop}%
\bibitem [{\citenamefont {Turner}\ \emph
  {et~al.}(2018{\natexlab{a}})\citenamefont {Turner}, \citenamefont
  {Michailidis}, \citenamefont {Abanin}, \citenamefont {Serbyn},\ and\
  \citenamefont {Papi{\'c}}}]{turner2018quantum}%
  \BibitemOpen
  \bibfield  {author} {\bibinfo {author} {\bibfnamefont {C.}~\bibnamefont
  {Turner}}, \bibinfo {author} {\bibfnamefont {A.}~\bibnamefont {Michailidis}},
  \bibinfo {author} {\bibfnamefont {D.}~\bibnamefont {Abanin}}, \bibinfo
  {author} {\bibfnamefont {M.}~\bibnamefont {Serbyn}}, \ and\ \bibinfo {author}
  {\bibfnamefont {Z.}~\bibnamefont {Papi{\'c}}},\ }\href@noop {} {\bibfield
  {journal} {\bibinfo  {journal} {Physical Review B}\ }\textbf {\bibinfo
  {volume} {98}},\ \bibinfo {pages} {155134} (\bibinfo {year}
  {2018}{\natexlab{a}})}\BibitemShut {NoStop}%
\bibitem [{\citenamefont {Turner}\ \emph
  {et~al.}(2018{\natexlab{b}})\citenamefont {Turner}, \citenamefont
  {Michailidis}, \citenamefont {Abanin}, \citenamefont {Serbyn},\ and\
  \citenamefont {Papi{\'c}}}]{turner2018weak}%
  \BibitemOpen
  \bibfield  {author} {\bibinfo {author} {\bibfnamefont {C.~J.}\ \bibnamefont
  {Turner}}, \bibinfo {author} {\bibfnamefont {A.~A.}\ \bibnamefont
  {Michailidis}}, \bibinfo {author} {\bibfnamefont {D.~A.}\ \bibnamefont
  {Abanin}}, \bibinfo {author} {\bibfnamefont {M.}~\bibnamefont {Serbyn}}, \
  and\ \bibinfo {author} {\bibfnamefont {Z.}~\bibnamefont {Papi{\'c}}},\
  }\href@noop {} {\bibfield  {journal} {\bibinfo  {journal} {Nature Physics}\
  }\textbf {\bibinfo {volume} {14}},\ \bibinfo {pages} {745} (\bibinfo {year}
  {2018}{\natexlab{b}})}\BibitemShut {NoStop}%
\bibitem [{\citenamefont {Serbyn}\ \emph {et~al.}(2021)\citenamefont {Serbyn},
  \citenamefont {Abanin},\ and\ \citenamefont {Papi{\'c}}}]{serbyn2021quantum}%
  \BibitemOpen
  \bibfield  {author} {\bibinfo {author} {\bibfnamefont {M.}~\bibnamefont
  {Serbyn}}, \bibinfo {author} {\bibfnamefont {D.~A.}\ \bibnamefont {Abanin}},
  \ and\ \bibinfo {author} {\bibfnamefont {Z.}~\bibnamefont {Papi{\'c}}},\
  }\href@noop {} {\bibfield  {journal} {\bibinfo  {journal} {Nature Physics}\
  }\textbf {\bibinfo {volume} {17}},\ \bibinfo {pages} {675} (\bibinfo {year}
  {2021})}\BibitemShut {NoStop}%
\bibitem [{\citenamefont {Chandran}\ \emph {et~al.}(2023)\citenamefont
  {Chandran}, \citenamefont {Iadecola}, \citenamefont {Khemani},\ and\
  \citenamefont {Moessner}}]{chandran2023quantum}%
  \BibitemOpen
  \bibfield  {author} {\bibinfo {author} {\bibfnamefont {A.}~\bibnamefont
  {Chandran}}, \bibinfo {author} {\bibfnamefont {T.}~\bibnamefont {Iadecola}},
  \bibinfo {author} {\bibfnamefont {V.}~\bibnamefont {Khemani}}, \ and\
  \bibinfo {author} {\bibfnamefont {R.}~\bibnamefont {Moessner}},\ }\href@noop
  {} {\bibfield  {journal} {\bibinfo  {journal} {Annual Review of Condensed
  Matter Physics}\ }\textbf {\bibinfo {volume} {14}},\ \bibinfo {pages} {443}
  (\bibinfo {year} {2023})}\BibitemShut {NoStop}%
\bibitem [{\citenamefont {Moudgalya}\ \emph {et~al.}(2022)\citenamefont
  {Moudgalya}, \citenamefont {Bernevig},\ and\ \citenamefont
  {Regnault}}]{moudgalya2022quantum}%
  \BibitemOpen
  \bibfield  {author} {\bibinfo {author} {\bibfnamefont {S.}~\bibnamefont
  {Moudgalya}}, \bibinfo {author} {\bibfnamefont {B.~A.}\ \bibnamefont
  {Bernevig}}, \ and\ \bibinfo {author} {\bibfnamefont {N.}~\bibnamefont
  {Regnault}},\ }\href@noop {} {\bibfield  {journal} {\bibinfo  {journal}
  {Reports on Progress in Physics}\ }\textbf {\bibinfo {volume} {85}},\
  \bibinfo {pages} {086501} (\bibinfo {year} {2022})}\BibitemShut {NoStop}%
\bibitem [{\citenamefont {Michailidis}\ \emph {et~al.}(2020)\citenamefont
  {Michailidis}, \citenamefont {Turner}, \citenamefont {Papi{\'c}},
  \citenamefont {Abanin},\ and\ \citenamefont {Serbyn}}]{michailidis2020slow}%
  \BibitemOpen
  \bibfield  {author} {\bibinfo {author} {\bibfnamefont {A.}~\bibnamefont
  {Michailidis}}, \bibinfo {author} {\bibfnamefont {C.}~\bibnamefont {Turner}},
  \bibinfo {author} {\bibfnamefont {Z.}~\bibnamefont {Papi{\'c}}}, \bibinfo
  {author} {\bibfnamefont {D.}~\bibnamefont {Abanin}}, \ and\ \bibinfo {author}
  {\bibfnamefont {M.}~\bibnamefont {Serbyn}},\ }\href@noop {} {\bibfield
  {journal} {\bibinfo  {journal} {Physical Review X}\ }\textbf {\bibinfo
  {volume} {10}},\ \bibinfo {pages} {011055} (\bibinfo {year}
  {2020})}\BibitemShut {NoStop}%
\bibitem [{\citenamefont {Nandy}\ \emph {et~al.}(2024)\citenamefont {Nandy},
  \citenamefont {Mukherjee}, \citenamefont {Bhattacharyya},\ and\ \citenamefont
  {Banerjee}}]{nandy2024quantum}%
  \BibitemOpen
  \bibfield  {author} {\bibinfo {author} {\bibfnamefont {S.}~\bibnamefont
  {Nandy}}, \bibinfo {author} {\bibfnamefont {B.}~\bibnamefont {Mukherjee}},
  \bibinfo {author} {\bibfnamefont {A.}~\bibnamefont {Bhattacharyya}}, \ and\
  \bibinfo {author} {\bibfnamefont {A.}~\bibnamefont {Banerjee}},\ }\href@noop
  {} {\bibfield  {journal} {\bibinfo  {journal} {Journal of Physics: Condensed
  Matter}\ }\textbf {\bibinfo {volume} {36}},\ \bibinfo {pages} {155601}
  (\bibinfo {year} {2024})}\BibitemShut {NoStop}%
\bibitem [{\citenamefont {Choi}\ \emph {et~al.}(2019)\citenamefont {Choi},
  \citenamefont {Turner}, \citenamefont {Pichler}, \citenamefont {Ho},
  \citenamefont {Michailidis}, \citenamefont {Papi{\'c}}, \citenamefont
  {Serbyn}, \citenamefont {Lukin},\ and\ \citenamefont
  {Abanin}}]{choi2019emergent}%
  \BibitemOpen
  \bibfield  {author} {\bibinfo {author} {\bibfnamefont {S.}~\bibnamefont
  {Choi}}, \bibinfo {author} {\bibfnamefont {C.~J.}\ \bibnamefont {Turner}},
  \bibinfo {author} {\bibfnamefont {H.}~\bibnamefont {Pichler}}, \bibinfo
  {author} {\bibfnamefont {W.~W.}\ \bibnamefont {Ho}}, \bibinfo {author}
  {\bibfnamefont {A.~A.}\ \bibnamefont {Michailidis}}, \bibinfo {author}
  {\bibfnamefont {Z.}~\bibnamefont {Papi{\'c}}}, \bibinfo {author}
  {\bibfnamefont {M.}~\bibnamefont {Serbyn}}, \bibinfo {author} {\bibfnamefont
  {M.~D.}\ \bibnamefont {Lukin}}, \ and\ \bibinfo {author} {\bibfnamefont
  {D.~A.}\ \bibnamefont {Abanin}},\ }\href@noop {} {\bibfield  {journal}
  {\bibinfo  {journal} {Physical review letters}\ }\textbf {\bibinfo {volume}
  {122}},\ \bibinfo {pages} {220603} (\bibinfo {year} {2019})}\BibitemShut
  {NoStop}%
\bibitem [{\citenamefont {Ho}\ \emph {et~al.}(2019)\citenamefont {Ho},
  \citenamefont {Choi}, \citenamefont {Pichler},\ and\ \citenamefont
  {Lukin}}]{ho2019periodic}%
  \BibitemOpen
  \bibfield  {author} {\bibinfo {author} {\bibfnamefont {W.~W.}\ \bibnamefont
  {Ho}}, \bibinfo {author} {\bibfnamefont {S.}~\bibnamefont {Choi}}, \bibinfo
  {author} {\bibfnamefont {H.}~\bibnamefont {Pichler}}, \ and\ \bibinfo
  {author} {\bibfnamefont {M.~D.}\ \bibnamefont {Lukin}},\ }\href@noop {}
  {\bibfield  {journal} {\bibinfo  {journal} {Physical Review Letters}\
  }\textbf {\bibinfo {volume} {122}},\ \bibinfo {pages} {040603} (\bibinfo
  {year} {2019})}\BibitemShut {NoStop}%
\bibitem [{\citenamefont {Zhao}\ \emph {et~al.}(2020)\citenamefont {Zhao},
  \citenamefont {Vovrosh}, \citenamefont {Mintert},\ and\ \citenamefont
  {Knolle}}]{zhao2020quantum}%
  \BibitemOpen
  \bibfield  {author} {\bibinfo {author} {\bibfnamefont {H.}~\bibnamefont
  {Zhao}}, \bibinfo {author} {\bibfnamefont {J.}~\bibnamefont {Vovrosh}},
  \bibinfo {author} {\bibfnamefont {F.}~\bibnamefont {Mintert}}, \ and\
  \bibinfo {author} {\bibfnamefont {J.}~\bibnamefont {Knolle}},\ }\href@noop {}
  {\bibfield  {journal} {\bibinfo  {journal} {Physical Review Letters}\
  }\textbf {\bibinfo {volume} {124}},\ \bibinfo {pages} {160604} (\bibinfo
  {year} {2020})}\BibitemShut {NoStop}%
\bibitem [{\citenamefont {Bull}\ \emph {et~al.}(2022)\citenamefont {Bull},
  \citenamefont {Hallam}, \citenamefont {Papi{\'c}},\ and\ \citenamefont
  {Martin}}]{bull2022tuning}%
  \BibitemOpen
  \bibfield  {author} {\bibinfo {author} {\bibfnamefont {K.}~\bibnamefont
  {Bull}}, \bibinfo {author} {\bibfnamefont {A.}~\bibnamefont {Hallam}},
  \bibinfo {author} {\bibfnamefont {Z.}~\bibnamefont {Papi{\'c}}}, \ and\
  \bibinfo {author} {\bibfnamefont {I.}~\bibnamefont {Martin}},\ }\href@noop {}
  {\bibfield  {journal} {\bibinfo  {journal} {Physical Review Letters}\
  }\textbf {\bibinfo {volume} {129}},\ \bibinfo {pages} {140602} (\bibinfo
  {year} {2022})}\BibitemShut {NoStop}%
\bibitem [{\citenamefont {Lin}\ and\ \citenamefont
  {Motrunich}(2019)}]{lin2019exact}%
  \BibitemOpen
  \bibfield  {author} {\bibinfo {author} {\bibfnamefont {C.-J.}\ \bibnamefont
  {Lin}}\ and\ \bibinfo {author} {\bibfnamefont {O.~I.}\ \bibnamefont
  {Motrunich}},\ }\href@noop {} {\bibfield  {journal} {\bibinfo  {journal}
  {Physical Review Letters}\ }\textbf {\bibinfo {volume} {122}},\ \bibinfo
  {pages} {173401} (\bibinfo {year} {2019})}\BibitemShut {NoStop}%
\bibitem [{\citenamefont {Mark}\ and\ \citenamefont
  {Motrunich}(2020)}]{mark2020eta}%
  \BibitemOpen
  \bibfield  {author} {\bibinfo {author} {\bibfnamefont {D.~K.}\ \bibnamefont
  {Mark}}\ and\ \bibinfo {author} {\bibfnamefont {O.~I.}\ \bibnamefont
  {Motrunich}},\ }\href@noop {} {\bibfield  {journal} {\bibinfo  {journal}
  {Physical Review B}\ }\textbf {\bibinfo {volume} {102}},\ \bibinfo {pages}
  {075132} (\bibinfo {year} {2020})}\BibitemShut {NoStop}%
\bibitem [{\citenamefont {Shibata}\ \emph {et~al.}(2020)\citenamefont
  {Shibata}, \citenamefont {Yoshioka},\ and\ \citenamefont
  {Katsura}}]{shibata2020onsager}%
  \BibitemOpen
  \bibfield  {author} {\bibinfo {author} {\bibfnamefont {N.}~\bibnamefont
  {Shibata}}, \bibinfo {author} {\bibfnamefont {N.}~\bibnamefont {Yoshioka}}, \
  and\ \bibinfo {author} {\bibfnamefont {H.}~\bibnamefont {Katsura}},\
  }\href@noop {} {\bibfield  {journal} {\bibinfo  {journal} {Physical Review
  Letters}\ }\textbf {\bibinfo {volume} {124}},\ \bibinfo {pages} {180604}
  (\bibinfo {year} {2020})}\BibitemShut {NoStop}%
\bibitem [{\citenamefont {Mohapatra}\ and\ \citenamefont
  {Balram}(2023)}]{mohapatra2023pronounced}%
  \BibitemOpen
  \bibfield  {author} {\bibinfo {author} {\bibfnamefont {S.}~\bibnamefont
  {Mohapatra}}\ and\ \bibinfo {author} {\bibfnamefont {A.~C.}\ \bibnamefont
  {Balram}},\ }\href@noop {} {\bibfield  {journal} {\bibinfo  {journal}
  {Physical Review B}\ }\textbf {\bibinfo {volume} {107}},\ \bibinfo {pages}
  {235121} (\bibinfo {year} {2023})}\BibitemShut {NoStop}%
\bibitem [{\citenamefont {Mondal}\ \emph {et~al.}(2020)\citenamefont {Mondal},
  \citenamefont {Sinha},\ and\ \citenamefont {Sinha}}]{mondal2020chaos}%
  \BibitemOpen
  \bibfield  {author} {\bibinfo {author} {\bibfnamefont {D.}~\bibnamefont
  {Mondal}}, \bibinfo {author} {\bibfnamefont {S.}~\bibnamefont {Sinha}}, \
  and\ \bibinfo {author} {\bibfnamefont {S.}~\bibnamefont {Sinha}},\
  }\href@noop {} {\bibfield  {journal} {\bibinfo  {journal} {Physical Review
  E}\ }\textbf {\bibinfo {volume} {102}},\ \bibinfo {pages} {020101} (\bibinfo
  {year} {2020})}\BibitemShut {NoStop}%
\bibitem [{\citenamefont {You}\ \emph {et~al.}(2022)\citenamefont {You},
  \citenamefont {Zhao}, \citenamefont {Ren}, \citenamefont {Sun}, \citenamefont
  {Li},\ and\ \citenamefont {Ole{\'s}}}]{you2022quantum}%
  \BibitemOpen
  \bibfield  {author} {\bibinfo {author} {\bibfnamefont {W.-L.}\ \bibnamefont
  {You}}, \bibinfo {author} {\bibfnamefont {Z.}~\bibnamefont {Zhao}}, \bibinfo
  {author} {\bibfnamefont {J.}~\bibnamefont {Ren}}, \bibinfo {author}
  {\bibfnamefont {G.}~\bibnamefont {Sun}}, \bibinfo {author} {\bibfnamefont
  {L.}~\bibnamefont {Li}}, \ and\ \bibinfo {author} {\bibfnamefont {A.~M.}\
  \bibnamefont {Ole{\'s}}},\ }\href@noop {} {\bibfield  {journal} {\bibinfo
  {journal} {Physical Review Research}\ }\textbf {\bibinfo {volume} {4}},\
  \bibinfo {pages} {013103} (\bibinfo {year} {2022})}\BibitemShut {NoStop}%
\bibitem [{\citenamefont {Schecter}\ and\ \citenamefont
  {Iadecola}(2019)}]{schecter2019weak}%
  \BibitemOpen
  \bibfield  {author} {\bibinfo {author} {\bibfnamefont {M.}~\bibnamefont
  {Schecter}}\ and\ \bibinfo {author} {\bibfnamefont {T.}~\bibnamefont
  {Iadecola}},\ }\href@noop {} {\bibfield  {journal} {\bibinfo  {journal}
  {Physical Review Letters}\ }\textbf {\bibinfo {volume} {123}},\ \bibinfo
  {pages} {147201} (\bibinfo {year} {2019})}\BibitemShut {NoStop}%
\bibitem [{\citenamefont {Lee}\ \emph {et~al.}(2020)\citenamefont {Lee},
  \citenamefont {Melendrez}, \citenamefont {Pal},\ and\ \citenamefont
  {Changlani}}]{lee2020exact}%
  \BibitemOpen
  \bibfield  {author} {\bibinfo {author} {\bibfnamefont {K.}~\bibnamefont
  {Lee}}, \bibinfo {author} {\bibfnamefont {R.}~\bibnamefont {Melendrez}},
  \bibinfo {author} {\bibfnamefont {A.}~\bibnamefont {Pal}}, \ and\ \bibinfo
  {author} {\bibfnamefont {H.~J.}\ \bibnamefont {Changlani}},\ }\href@noop {}
  {\bibfield  {journal} {\bibinfo  {journal} {Physical Review B}\ }\textbf
  {\bibinfo {volume} {101}},\ \bibinfo {pages} {241111} (\bibinfo {year}
  {2020})}\BibitemShut {NoStop}%
\bibitem [{\citenamefont {McClarty}\ \emph {et~al.}(2020)\citenamefont
  {McClarty}, \citenamefont {Haque}, \citenamefont {Sen},\ and\ \citenamefont
  {Richter}}]{mcclarty2020disorder}%
  \BibitemOpen
  \bibfield  {author} {\bibinfo {author} {\bibfnamefont {P.~A.}\ \bibnamefont
  {McClarty}}, \bibinfo {author} {\bibfnamefont {M.}~\bibnamefont {Haque}},
  \bibinfo {author} {\bibfnamefont {A.}~\bibnamefont {Sen}}, \ and\ \bibinfo
  {author} {\bibfnamefont {J.}~\bibnamefont {Richter}},\ }\href@noop {}
  {\bibfield  {journal} {\bibinfo  {journal} {Physical Review B}\ }\textbf
  {\bibinfo {volume} {102}},\ \bibinfo {pages} {224303} (\bibinfo {year}
  {2020})}\BibitemShut {NoStop}%
\bibitem [{\citenamefont {Khare}\ and\ \citenamefont
  {Choudhury}(2020)}]{khare2020localized}%
  \BibitemOpen
  \bibfield  {author} {\bibinfo {author} {\bibfnamefont {R.}~\bibnamefont
  {Khare}}\ and\ \bibinfo {author} {\bibfnamefont {S.}~\bibnamefont
  {Choudhury}},\ }\href@noop {} {\bibfield  {journal} {\bibinfo  {journal}
  {Journal of Physics B: Atomic, Molecular and Optical Physics}\ }\textbf
  {\bibinfo {volume} {54}},\ \bibinfo {pages} {015301} (\bibinfo {year}
  {2020})}\BibitemShut {NoStop}%
\bibitem [{\citenamefont {Chertkov}\ and\ \citenamefont
  {Clark}(2021)}]{chertkov2021motif}%
  \BibitemOpen
  \bibfield  {author} {\bibinfo {author} {\bibfnamefont {E.}~\bibnamefont
  {Chertkov}}\ and\ \bibinfo {author} {\bibfnamefont {B.~K.}\ \bibnamefont
  {Clark}},\ }\href@noop {} {\bibfield  {journal} {\bibinfo  {journal}
  {Physical Review B}\ }\textbf {\bibinfo {volume} {104}},\ \bibinfo {pages}
  {104410} (\bibinfo {year} {2021})}\BibitemShut {NoStop}%
\bibitem [{\citenamefont {Banerjee}\ and\ \citenamefont
  {Sen}(2021)}]{banerjee2021quantum}%
  \BibitemOpen
  \bibfield  {author} {\bibinfo {author} {\bibfnamefont {D.}~\bibnamefont
  {Banerjee}}\ and\ \bibinfo {author} {\bibfnamefont {A.}~\bibnamefont {Sen}},\
  }\href@noop {} {\bibfield  {journal} {\bibinfo  {journal} {Physical Review
  Letters}\ }\textbf {\bibinfo {volume} {126}},\ \bibinfo {pages} {220601}
  (\bibinfo {year} {2021})}\BibitemShut {NoStop}%
\bibitem [{\citenamefont {Udupa}\ \emph {et~al.}(2023)\citenamefont {Udupa},
  \citenamefont {Sur}, \citenamefont {Nandy}, \citenamefont {Sen},\ and\
  \citenamefont {Sen}}]{udupa2023weak}%
  \BibitemOpen
  \bibfield  {author} {\bibinfo {author} {\bibfnamefont {A.}~\bibnamefont
  {Udupa}}, \bibinfo {author} {\bibfnamefont {S.}~\bibnamefont {Sur}}, \bibinfo
  {author} {\bibfnamefont {S.}~\bibnamefont {Nandy}}, \bibinfo {author}
  {\bibfnamefont {A.}~\bibnamefont {Sen}}, \ and\ \bibinfo {author}
  {\bibfnamefont {D.}~\bibnamefont {Sen}},\ }\href@noop {} {\bibfield
  {journal} {\bibinfo  {journal} {Physical Review B}\ }\textbf {\bibinfo
  {volume} {108}},\ \bibinfo {pages} {214430} (\bibinfo {year}
  {2023})}\BibitemShut {NoStop}%
\bibitem [{\citenamefont {Biswas}\ \emph {et~al.}(2022)\citenamefont {Biswas},
  \citenamefont {Banerjee},\ and\ \citenamefont {Sen}}]{biswas2022scars}%
  \BibitemOpen
  \bibfield  {author} {\bibinfo {author} {\bibfnamefont {S.}~\bibnamefont
  {Biswas}}, \bibinfo {author} {\bibfnamefont {D.}~\bibnamefont {Banerjee}}, \
  and\ \bibinfo {author} {\bibfnamefont {A.}~\bibnamefont {Sen}},\ }\href@noop
  {} {\bibfield  {journal} {\bibinfo  {journal} {SciPost Physics}\ }\textbf
  {\bibinfo {volume} {12}},\ \bibinfo {pages} {148} (\bibinfo {year}
  {2022})}\BibitemShut {NoStop}%
\bibitem [{\citenamefont {Desaules}\ \emph {et~al.}(2023)\citenamefont
  {Desaules}, \citenamefont {Banerjee}, \citenamefont {Hudomal}, \citenamefont
  {Papi{\'c}}, \citenamefont {Sen},\ and\ \citenamefont
  {Halimeh}}]{desaules2023weak}%
  \BibitemOpen
  \bibfield  {author} {\bibinfo {author} {\bibfnamefont {J.-Y.}\ \bibnamefont
  {Desaules}}, \bibinfo {author} {\bibfnamefont {D.}~\bibnamefont {Banerjee}},
  \bibinfo {author} {\bibfnamefont {A.}~\bibnamefont {Hudomal}}, \bibinfo
  {author} {\bibfnamefont {Z.}~\bibnamefont {Papi{\'c}}}, \bibinfo {author}
  {\bibfnamefont {A.}~\bibnamefont {Sen}}, \ and\ \bibinfo {author}
  {\bibfnamefont {J.~C.}\ \bibnamefont {Halimeh}},\ }\href@noop {} {\bibfield
  {journal} {\bibinfo  {journal} {Physical Review B}\ }\textbf {\bibinfo
  {volume} {107}},\ \bibinfo {pages} {L201105} (\bibinfo {year}
  {2023})}\BibitemShut {NoStop}%
\bibitem [{\citenamefont {Mukherjee}\ \emph {et~al.}(2021)\citenamefont
  {Mukherjee}, \citenamefont {Cai},\ and\ \citenamefont
  {Liu}}]{mukherjee2021constraint}%
  \BibitemOpen
  \bibfield  {author} {\bibinfo {author} {\bibfnamefont {B.}~\bibnamefont
  {Mukherjee}}, \bibinfo {author} {\bibfnamefont {Z.}~\bibnamefont {Cai}}, \
  and\ \bibinfo {author} {\bibfnamefont {W.~V.}\ \bibnamefont {Liu}},\
  }\href@noop {} {\bibfield  {journal} {\bibinfo  {journal} {Physical Review
  Research}\ }\textbf {\bibinfo {volume} {3}},\ \bibinfo {pages} {033201}
  (\bibinfo {year} {2021})}\BibitemShut {NoStop}%
\bibitem [{\citenamefont {Chen}\ and\ \citenamefont
  {Cai}(2020)}]{chen2020persistent}%
  \BibitemOpen
  \bibfield  {author} {\bibinfo {author} {\bibfnamefont {Y.}~\bibnamefont
  {Chen}}\ and\ \bibinfo {author} {\bibfnamefont {Z.}~\bibnamefont {Cai}},\
  }\href@noop {} {\bibfield  {journal} {\bibinfo  {journal} {Physical Review
  A}\ }\textbf {\bibinfo {volume} {101}},\ \bibinfo {pages} {023611} (\bibinfo
  {year} {2020})}\BibitemShut {NoStop}%
\bibitem [{\citenamefont {Lesanovsky}\ and\ \citenamefont
  {Katsura}(2012)}]{lesanovsky2012interacting}%
  \BibitemOpen
  \bibfield  {author} {\bibinfo {author} {\bibfnamefont {I.}~\bibnamefont
  {Lesanovsky}}\ and\ \bibinfo {author} {\bibfnamefont {H.}~\bibnamefont
  {Katsura}},\ }\href@noop {} {\bibfield  {journal} {\bibinfo  {journal}
  {Physical Review A}\ }\textbf {\bibinfo {volume} {86}},\ \bibinfo {pages}
  {041601} (\bibinfo {year} {2012})}\BibitemShut {NoStop}%
\bibitem [{\citenamefont {Mukherjee}\ \emph
  {et~al.}(2020{\natexlab{a}})\citenamefont {Mukherjee}, \citenamefont {Nandy},
  \citenamefont {Sen}, \citenamefont {Sen},\ and\ \citenamefont
  {Sengupta}}]{mukherjee2020collapse}%
  \BibitemOpen
  \bibfield  {author} {\bibinfo {author} {\bibfnamefont {B.}~\bibnamefont
  {Mukherjee}}, \bibinfo {author} {\bibfnamefont {S.}~\bibnamefont {Nandy}},
  \bibinfo {author} {\bibfnamefont {A.}~\bibnamefont {Sen}}, \bibinfo {author}
  {\bibfnamefont {D.}~\bibnamefont {Sen}}, \ and\ \bibinfo {author}
  {\bibfnamefont {K.}~\bibnamefont {Sengupta}},\ }\href@noop {} {\bibfield
  {journal} {\bibinfo  {journal} {Physical Review B}\ }\textbf {\bibinfo
  {volume} {101}},\ \bibinfo {pages} {245107} (\bibinfo {year}
  {2020}{\natexlab{a}})}\BibitemShut {NoStop}%
\bibitem [{\citenamefont {Mukherjee}\ \emph
  {et~al.}(2020{\natexlab{b}})\citenamefont {Mukherjee}, \citenamefont {Sen},
  \citenamefont {Sen},\ and\ \citenamefont
  {Sengupta}}]{mukherjee2020restoring}%
  \BibitemOpen
  \bibfield  {author} {\bibinfo {author} {\bibfnamefont {B.}~\bibnamefont
  {Mukherjee}}, \bibinfo {author} {\bibfnamefont {A.}~\bibnamefont {Sen}},
  \bibinfo {author} {\bibfnamefont {D.}~\bibnamefont {Sen}}, \ and\ \bibinfo
  {author} {\bibfnamefont {K.}~\bibnamefont {Sengupta}},\ }\href@noop {}
  {\bibfield  {journal} {\bibinfo  {journal} {Physical Review B}\ }\textbf
  {\bibinfo {volume} {102}},\ \bibinfo {pages} {014301} (\bibinfo {year}
  {2020}{\natexlab{b}})}\BibitemShut {NoStop}%
\bibitem [{\citenamefont {Mukherjee}\ \emph {et~al.}(2022)\citenamefont
  {Mukherjee}, \citenamefont {Sen},\ and\ \citenamefont
  {Sengupta}}]{mukherjee2022periodically}%
  \BibitemOpen
  \bibfield  {author} {\bibinfo {author} {\bibfnamefont {B.}~\bibnamefont
  {Mukherjee}}, \bibinfo {author} {\bibfnamefont {A.}~\bibnamefont {Sen}}, \
  and\ \bibinfo {author} {\bibfnamefont {K.}~\bibnamefont {Sengupta}},\
  }\href@noop {} {\bibfield  {journal} {\bibinfo  {journal} {Physical Review
  B}\ }\textbf {\bibinfo {volume} {106}},\ \bibinfo {pages} {064305} (\bibinfo
  {year} {2022})}\BibitemShut {NoStop}%
\bibitem [{\citenamefont {Banerjee}\ \emph {et~al.}(2024)\citenamefont
  {Banerjee}, \citenamefont {Choudhury},\ and\ \citenamefont
  {Sengupta}}]{banerjee2024exact}%
  \BibitemOpen
  \bibfield  {author} {\bibinfo {author} {\bibfnamefont {T.}~\bibnamefont
  {Banerjee}}, \bibinfo {author} {\bibfnamefont {S.}~\bibnamefont {Choudhury}},
  \ and\ \bibinfo {author} {\bibfnamefont {K.}~\bibnamefont {Sengupta}},\
  }\href@noop {} {\bibfield  {journal} {\bibinfo  {journal} {arXiv preprint
  arXiv:2404.06536}\ } (\bibinfo {year} {2024})}\BibitemShut {NoStop}%
\bibitem [{\citenamefont {Deng}\ and\ \citenamefont
  {Yang}(2023)}]{deng2023using}%
  \BibitemOpen
  \bibfield  {author} {\bibinfo {author} {\bibfnamefont {W.}~\bibnamefont
  {Deng}}\ and\ \bibinfo {author} {\bibfnamefont {Z.-C.}\ \bibnamefont
  {Yang}},\ }\href@noop {} {\bibfield  {journal} {\bibinfo  {journal} {Physical
  Review B}\ }\textbf {\bibinfo {volume} {108}},\ \bibinfo {pages} {205129}
  (\bibinfo {year} {2023})}\BibitemShut {NoStop}%
\bibitem [{\citenamefont {Huang}\ and\ \citenamefont
  {Li}(2024)}]{huang2024engineering}%
  \BibitemOpen
  \bibfield  {author} {\bibinfo {author} {\bibfnamefont {K.}~\bibnamefont
  {Huang}}\ and\ \bibinfo {author} {\bibfnamefont {X.}~\bibnamefont {Li}},\
  }\href@noop {} {\bibfield  {journal} {\bibinfo  {journal} {Physical Review
  B}\ }\textbf {\bibinfo {volume} {109}},\ \bibinfo {pages} {064306} (\bibinfo
  {year} {2024})}\BibitemShut {NoStop}%
\bibitem [{\citenamefont {Maskara}\ \emph {et~al.}(2021)\citenamefont
  {Maskara}, \citenamefont {Michailidis}, \citenamefont {Ho}, \citenamefont
  {Bluvstein}, \citenamefont {Choi}, \citenamefont {Lukin},\ and\ \citenamefont
  {Serbyn}}]{maskara2021discrete}%
  \BibitemOpen
  \bibfield  {author} {\bibinfo {author} {\bibfnamefont {N.}~\bibnamefont
  {Maskara}}, \bibinfo {author} {\bibfnamefont {A.~A.}\ \bibnamefont
  {Michailidis}}, \bibinfo {author} {\bibfnamefont {W.~W.}\ \bibnamefont {Ho}},
  \bibinfo {author} {\bibfnamefont {D.}~\bibnamefont {Bluvstein}}, \bibinfo
  {author} {\bibfnamefont {S.}~\bibnamefont {Choi}}, \bibinfo {author}
  {\bibfnamefont {M.~D.}\ \bibnamefont {Lukin}}, \ and\ \bibinfo {author}
  {\bibfnamefont {M.}~\bibnamefont {Serbyn}},\ }\href@noop {} {\bibfield
  {journal} {\bibinfo  {journal} {Physical Review Letters}\ }\textbf {\bibinfo
  {volume} {127}},\ \bibinfo {pages} {090602} (\bibinfo {year}
  {2021})}\BibitemShut {NoStop}%
\bibitem [{\citenamefont {Bluvstein}\ \emph {et~al.}(2021)\citenamefont
  {Bluvstein}, \citenamefont {Omran}, \citenamefont {Levine}, \citenamefont
  {Keesling}, \citenamefont {Semeghini}, \citenamefont {Ebadi}, \citenamefont
  {Wang}, \citenamefont {Michailidis}, \citenamefont {Maskara}, \citenamefont
  {Ho} \emph {et~al.}}]{bluvstein2021controlling}%
  \BibitemOpen
  \bibfield  {author} {\bibinfo {author} {\bibfnamefont {D.}~\bibnamefont
  {Bluvstein}}, \bibinfo {author} {\bibfnamefont {A.}~\bibnamefont {Omran}},
  \bibinfo {author} {\bibfnamefont {H.}~\bibnamefont {Levine}}, \bibinfo
  {author} {\bibfnamefont {A.}~\bibnamefont {Keesling}}, \bibinfo {author}
  {\bibfnamefont {G.}~\bibnamefont {Semeghini}}, \bibinfo {author}
  {\bibfnamefont {S.}~\bibnamefont {Ebadi}}, \bibinfo {author} {\bibfnamefont
  {T.~T.}\ \bibnamefont {Wang}}, \bibinfo {author} {\bibfnamefont {A.~A.}\
  \bibnamefont {Michailidis}}, \bibinfo {author} {\bibfnamefont
  {N.}~\bibnamefont {Maskara}}, \bibinfo {author} {\bibfnamefont {W.~W.}\
  \bibnamefont {Ho}},  \emph {et~al.},\ }\href@noop {} {\bibfield  {journal}
  {\bibinfo  {journal} {Science}\ }\textbf {\bibinfo {volume} {371}},\ \bibinfo
  {pages} {1355} (\bibinfo {year} {2021})}\BibitemShut {NoStop}%
\bibitem [{\citenamefont {Park}\ and\ \citenamefont
  {Lee}(2023)}]{park2023subharmonic}%
  \BibitemOpen
  \bibfield  {author} {\bibinfo {author} {\bibfnamefont {H.~K.}\ \bibnamefont
  {Park}}\ and\ \bibinfo {author} {\bibfnamefont {S.}~\bibnamefont {Lee}},\
  }\href@noop {} {\bibfield  {journal} {\bibinfo  {journal} {Physical Review
  B}\ }\textbf {\bibinfo {volume} {107}},\ \bibinfo {pages} {205142} (\bibinfo
  {year} {2023})}\BibitemShut {NoStop}%
\bibitem [{\citenamefont {Else}\ \emph
  {et~al.}(2020{\natexlab{b}})\citenamefont {Else}, \citenamefont {Ho},\ and\
  \citenamefont {Dumitrescu}}]{else2020long}%
  \BibitemOpen
  \bibfield  {author} {\bibinfo {author} {\bibfnamefont {D.~V.}\ \bibnamefont
  {Else}}, \bibinfo {author} {\bibfnamefont {W.~W.}\ \bibnamefont {Ho}}, \ and\
  \bibinfo {author} {\bibfnamefont {P.~T.}\ \bibnamefont {Dumitrescu}},\
  }\href@noop {} {\bibfield  {journal} {\bibinfo  {journal} {Physical Review
  X}\ }\textbf {\bibinfo {volume} {10}},\ \bibinfo {pages} {021032} (\bibinfo
  {year} {2020}{\natexlab{b}})}\BibitemShut {NoStop}%
\bibitem [{\citenamefont {He}\ \emph {et~al.}(2024)\citenamefont {He},
  \citenamefont {Ye}, \citenamefont {Gong}, \citenamefont {Yao}, \citenamefont
  {Liu}, \citenamefont {Murch}, \citenamefont {Yao},\ and\ \citenamefont
  {Zu}}]{he2024experimental}%
  \BibitemOpen
  \bibfield  {author} {\bibinfo {author} {\bibfnamefont {G.}~\bibnamefont
  {He}}, \bibinfo {author} {\bibfnamefont {B.}~\bibnamefont {Ye}}, \bibinfo
  {author} {\bibfnamefont {R.}~\bibnamefont {Gong}}, \bibinfo {author}
  {\bibfnamefont {C.}~\bibnamefont {Yao}}, \bibinfo {author} {\bibfnamefont
  {Z.}~\bibnamefont {Liu}}, \bibinfo {author} {\bibfnamefont {K.~W.}\
  \bibnamefont {Murch}}, \bibinfo {author} {\bibfnamefont {N.~Y.}\ \bibnamefont
  {Yao}}, \ and\ \bibinfo {author} {\bibfnamefont {C.}~\bibnamefont {Zu}},\
  }\href@noop {} {\bibfield  {journal} {\bibinfo  {journal} {arXiv preprint
  arXiv:2403.17842}\ } (\bibinfo {year} {2024})}\BibitemShut {NoStop}%
\bibitem [{\citenamefont {Zhao}\ \emph {et~al.}(2019)\citenamefont {Zhao},
  \citenamefont {Mintert},\ and\ \citenamefont {Knolle}}]{zhao2019floquet}%
  \BibitemOpen
  \bibfield  {author} {\bibinfo {author} {\bibfnamefont {H.}~\bibnamefont
  {Zhao}}, \bibinfo {author} {\bibfnamefont {F.}~\bibnamefont {Mintert}}, \
  and\ \bibinfo {author} {\bibfnamefont {J.}~\bibnamefont {Knolle}},\
  }\href@noop {} {\bibfield  {journal} {\bibinfo  {journal} {Physical Review
  B}\ }\textbf {\bibinfo {volume} {100}},\ \bibinfo {pages} {134302} (\bibinfo
  {year} {2019})}\BibitemShut {NoStop}%
\bibitem [{\citenamefont {Zhao}\ \emph {et~al.}(2023)\citenamefont {Zhao},
  \citenamefont {Knolle},\ and\ \citenamefont {Moessner}}]{zhao2023temporal}%
  \BibitemOpen
  \bibfield  {author} {\bibinfo {author} {\bibfnamefont {H.}~\bibnamefont
  {Zhao}}, \bibinfo {author} {\bibfnamefont {J.}~\bibnamefont {Knolle}}, \ and\
  \bibinfo {author} {\bibfnamefont {R.}~\bibnamefont {Moessner}},\ }\href@noop
  {} {\bibfield  {journal} {\bibinfo  {journal} {Physical Review B}\ }\textbf
  {\bibinfo {volume} {108}},\ \bibinfo {pages} {L100203} (\bibinfo {year}
  {2023})}\BibitemShut {NoStop}%
\bibitem [{\citenamefont {Moon}\ \emph {et~al.}(2024)\citenamefont {Moon},
  \citenamefont {Schindler}, \citenamefont {Sun}, \citenamefont {Druga},
  \citenamefont {Knolle}, \citenamefont {Moessner}, \citenamefont {Zhao},
  \citenamefont {Bukov},\ and\ \citenamefont {Ajoy}}]{moon2024experimental}%
  \BibitemOpen
  \bibfield  {author} {\bibinfo {author} {\bibfnamefont {L.~J.~I.}\
  \bibnamefont {Moon}}, \bibinfo {author} {\bibfnamefont {P.~M.}\ \bibnamefont
  {Schindler}}, \bibinfo {author} {\bibfnamefont {Y.}~\bibnamefont {Sun}},
  \bibinfo {author} {\bibfnamefont {E.}~\bibnamefont {Druga}}, \bibinfo
  {author} {\bibfnamefont {J.}~\bibnamefont {Knolle}}, \bibinfo {author}
  {\bibfnamefont {R.}~\bibnamefont {Moessner}}, \bibinfo {author}
  {\bibfnamefont {H.}~\bibnamefont {Zhao}}, \bibinfo {author} {\bibfnamefont
  {M.}~\bibnamefont {Bukov}}, \ and\ \bibinfo {author} {\bibfnamefont
  {A.}~\bibnamefont {Ajoy}},\ }\href@noop {} {\bibfield  {journal} {\bibinfo
  {journal} {arXiv preprint arXiv:2404.05620}\ } (\bibinfo {year}
  {2024})}\BibitemShut {NoStop}%
\bibitem [{\citenamefont {Choudhury}\ and\ \citenamefont
  {Liu}(2021)}]{choudhury2021self}%
  \BibitemOpen
  \bibfield  {author} {\bibinfo {author} {\bibfnamefont {S.}~\bibnamefont
  {Choudhury}}\ and\ \bibinfo {author} {\bibfnamefont {W.~V.}\ \bibnamefont
  {Liu}},\ }\href@noop {} {\bibfield  {journal} {\bibinfo  {journal} {arXiv
  preprint arXiv:2109.05318}\ } (\bibinfo {year} {2021})}\BibitemShut {NoStop}%
\bibitem [{\citenamefont {Kumar}\ and\ \citenamefont
  {Choudhury}(2024)}]{kumar2024prethermalization}%
  \BibitemOpen
  \bibfield  {author} {\bibinfo {author} {\bibfnamefont {S.}~\bibnamefont
  {Kumar}}\ and\ \bibinfo {author} {\bibfnamefont {S.}~\bibnamefont
  {Choudhury}},\ }\href@noop {} {\bibfield  {journal} {\bibinfo  {journal}
  {arXiv preprint arXiv:2404.10224}\ } (\bibinfo {year} {2024})}\BibitemShut
  {NoStop}%
\bibitem [{\citenamefont {Mori}\ \emph {et~al.}(2021)\citenamefont {Mori},
  \citenamefont {Zhao}, \citenamefont {Mintert}, \citenamefont {Knolle},\ and\
  \citenamefont {Moessner}}]{mori2021rigorous}%
  \BibitemOpen
  \bibfield  {author} {\bibinfo {author} {\bibfnamefont {T.}~\bibnamefont
  {Mori}}, \bibinfo {author} {\bibfnamefont {H.}~\bibnamefont {Zhao}}, \bibinfo
  {author} {\bibfnamefont {F.}~\bibnamefont {Mintert}}, \bibinfo {author}
  {\bibfnamefont {J.}~\bibnamefont {Knolle}}, \ and\ \bibinfo {author}
  {\bibfnamefont {R.}~\bibnamefont {Moessner}},\ }\href@noop {} {\bibfield
  {journal} {\bibinfo  {journal} {Physical Review Letters}\ }\textbf {\bibinfo
  {volume} {127}},\ \bibinfo {pages} {050602} (\bibinfo {year}
  {2021})}\BibitemShut {NoStop}%
\bibitem [{\citenamefont {Pilatowsky-Cameo}\ \emph {et~al.}(2023)\citenamefont
  {Pilatowsky-Cameo}, \citenamefont {Dag}, \citenamefont {Ho},\ and\
  \citenamefont {Choi}}]{pilatowsky2023complete}%
  \BibitemOpen
  \bibfield  {author} {\bibinfo {author} {\bibfnamefont {S.}~\bibnamefont
  {Pilatowsky-Cameo}}, \bibinfo {author} {\bibfnamefont {C.~B.}\ \bibnamefont
  {Dag}}, \bibinfo {author} {\bibfnamefont {W.~W.}\ \bibnamefont {Ho}}, \ and\
  \bibinfo {author} {\bibfnamefont {S.}~\bibnamefont {Choi}},\ }\href@noop {}
  {\bibfield  {journal} {\bibinfo  {journal} {Physical Review Letters}\
  }\textbf {\bibinfo {volume} {131}},\ \bibinfo {pages} {250401} (\bibinfo
  {year} {2023})}\BibitemShut {NoStop}%
\bibitem [{\citenamefont {Yan}\ \emph {et~al.}(2024)\citenamefont {Yan},
  \citenamefont {Moessner},\ and\ \citenamefont
  {Zhao}}]{yan2024prethermalization}%
  \BibitemOpen
  \bibfield  {author} {\bibinfo {author} {\bibfnamefont {J.}~\bibnamefont
  {Yan}}, \bibinfo {author} {\bibfnamefont {R.}~\bibnamefont {Moessner}}, \
  and\ \bibinfo {author} {\bibfnamefont {H.}~\bibnamefont {Zhao}},\ }\href@noop
  {} {\bibfield  {journal} {\bibinfo  {journal} {Physical Review B}\ }\textbf
  {\bibinfo {volume} {109}},\ \bibinfo {pages} {064305} (\bibinfo {year}
  {2024})}\BibitemShut {NoStop}%
\bibitem [{\citenamefont {Cai}(2022)}]{cai20221}%
  \BibitemOpen
  \bibfield  {author} {\bibinfo {author} {\bibfnamefont {Z.}~\bibnamefont
  {Cai}},\ }\href@noop {} {\bibfield  {journal} {\bibinfo  {journal} {Physical
  Review Letters}\ }\textbf {\bibinfo {volume} {128}},\ \bibinfo {pages}
  {050601} (\bibinfo {year} {2022})}\BibitemShut {NoStop}%
\bibitem [{\citenamefont {Nandy}\ \emph {et~al.}(2017)\citenamefont {Nandy},
  \citenamefont {Sen},\ and\ \citenamefont {Sen}}]{nandy2017aperiodically}%
  \BibitemOpen
  \bibfield  {author} {\bibinfo {author} {\bibfnamefont {S.}~\bibnamefont
  {Nandy}}, \bibinfo {author} {\bibfnamefont {A.}~\bibnamefont {Sen}}, \ and\
  \bibinfo {author} {\bibfnamefont {D.}~\bibnamefont {Sen}},\ }\href@noop {}
  {\bibfield  {journal} {\bibinfo  {journal} {Physical Review X}\ }\textbf
  {\bibinfo {volume} {7}},\ \bibinfo {pages} {031034} (\bibinfo {year}
  {2017})}\BibitemShut {NoStop}%
\bibitem [{\citenamefont {Zhao}\ \emph {et~al.}(2021)\citenamefont {Zhao},
  \citenamefont {Mintert}, \citenamefont {Moessner},\ and\ \citenamefont
  {Knolle}}]{zhao2021random}%
  \BibitemOpen
  \bibfield  {author} {\bibinfo {author} {\bibfnamefont {H.}~\bibnamefont
  {Zhao}}, \bibinfo {author} {\bibfnamefont {F.}~\bibnamefont {Mintert}},
  \bibinfo {author} {\bibfnamefont {R.}~\bibnamefont {Moessner}}, \ and\
  \bibinfo {author} {\bibfnamefont {J.}~\bibnamefont {Knolle}},\ }\href@noop {}
  {\bibfield  {journal} {\bibinfo  {journal} {Physical Review Letters}\
  }\textbf {\bibinfo {volume} {126}},\ \bibinfo {pages} {040601} (\bibinfo
  {year} {2021})}\BibitemShut {NoStop}%
\bibitem [{\citenamefont {Zhao}\ \emph
  {et~al.}(2022{\natexlab{a}})\citenamefont {Zhao}, \citenamefont {Mintert},
  \citenamefont {Knolle},\ and\ \citenamefont
  {Moessner}}]{zhao2022localization}%
  \BibitemOpen
  \bibfield  {author} {\bibinfo {author} {\bibfnamefont {H.}~\bibnamefont
  {Zhao}}, \bibinfo {author} {\bibfnamefont {F.}~\bibnamefont {Mintert}},
  \bibinfo {author} {\bibfnamefont {J.}~\bibnamefont {Knolle}}, \ and\ \bibinfo
  {author} {\bibfnamefont {R.}~\bibnamefont {Moessner}},\ }\href@noop {}
  {\bibfield  {journal} {\bibinfo  {journal} {Physical Review B}\ }\textbf
  {\bibinfo {volume} {105}},\ \bibinfo {pages} {L220202} (\bibinfo {year}
  {2022}{\natexlab{a}})}\BibitemShut {NoStop}%
\bibitem [{\citenamefont {Zhao}\ \emph
  {et~al.}(2022{\natexlab{b}})\citenamefont {Zhao}, \citenamefont {Knolle},
  \citenamefont {Moessner},\ and\ \citenamefont
  {Mintert}}]{zhao2022suppression}%
  \BibitemOpen
  \bibfield  {author} {\bibinfo {author} {\bibfnamefont {H.}~\bibnamefont
  {Zhao}}, \bibinfo {author} {\bibfnamefont {J.}~\bibnamefont {Knolle}},
  \bibinfo {author} {\bibfnamefont {R.}~\bibnamefont {Moessner}}, \ and\
  \bibinfo {author} {\bibfnamefont {F.}~\bibnamefont {Mintert}},\ }\href@noop
  {} {\bibfield  {journal} {\bibinfo  {journal} {Physical Review Letters}\
  }\textbf {\bibinfo {volume} {129}},\ \bibinfo {pages} {120605} (\bibinfo
  {year} {2022}{\natexlab{b}})}\BibitemShut {NoStop}%
\bibitem [{\citenamefont {Semeghini}\ \emph {et~al.}(2021)\citenamefont
  {Semeghini}, \citenamefont {Levine}, \citenamefont {Keesling}, \citenamefont
  {Ebadi}, \citenamefont {Wang}, \citenamefont {Bluvstein}, \citenamefont
  {Verresen}, \citenamefont {Pichler}, \citenamefont {Kalinowski},
  \citenamefont {Samajdar}, \citenamefont {Omran}, \citenamefont {Sachdev},
  \citenamefont {Vishwanath}, \citenamefont {Greiner}, \citenamefont
  {Vuleti\'{c}},\ and\ \citenamefont {Lukin}}]{semeghini2021probing}%
  \BibitemOpen
  \bibfield  {author} {\bibinfo {author} {\bibfnamefont {G.}~\bibnamefont
  {Semeghini}}, \bibinfo {author} {\bibfnamefont {H.}~\bibnamefont {Levine}},
  \bibinfo {author} {\bibfnamefont {A.}~\bibnamefont {Keesling}}, \bibinfo
  {author} {\bibfnamefont {S.}~\bibnamefont {Ebadi}}, \bibinfo {author}
  {\bibfnamefont {T.~T.}\ \bibnamefont {Wang}}, \bibinfo {author}
  {\bibfnamefont {D.}~\bibnamefont {Bluvstein}}, \bibinfo {author}
  {\bibfnamefont {R.}~\bibnamefont {Verresen}}, \bibinfo {author}
  {\bibfnamefont {H.}~\bibnamefont {Pichler}}, \bibinfo {author} {\bibfnamefont
  {M.}~\bibnamefont {Kalinowski}}, \bibinfo {author} {\bibfnamefont
  {R.}~\bibnamefont {Samajdar}}, \bibinfo {author} {\bibfnamefont
  {A.}~\bibnamefont {Omran}}, \bibinfo {author} {\bibfnamefont
  {S.}~\bibnamefont {Sachdev}}, \bibinfo {author} {\bibfnamefont
  {A.}~\bibnamefont {Vishwanath}}, \bibinfo {author} {\bibfnamefont
  {M.}~\bibnamefont {Greiner}}, \bibinfo {author} {\bibfnamefont
  {V.}~\bibnamefont {Vuleti\'{c}}}, \ and\ \bibinfo {author} {\bibfnamefont
  {M.~D.}\ \bibnamefont {Lukin}},\ }\href@noop {} {\bibfield  {journal}
  {\bibinfo  {journal} {Science}\ }\textbf {\bibinfo {volume} {374}},\ \bibinfo
  {pages} {1242} (\bibinfo {year} {2021})}\BibitemShut {NoStop}%
\bibitem [{\citenamefont {Verresen}\ \emph {et~al.}(2021)\citenamefont
  {Verresen}, \citenamefont {Lukin},\ and\ \citenamefont
  {Vishwanath}}]{verresen2021prediction}%
  \BibitemOpen
  \bibfield  {author} {\bibinfo {author} {\bibfnamefont {R.}~\bibnamefont
  {Verresen}}, \bibinfo {author} {\bibfnamefont {M.~D.}\ \bibnamefont {Lukin}},
  \ and\ \bibinfo {author} {\bibfnamefont {A.}~\bibnamefont {Vishwanath}},\
  }\href@noop {} {\bibfield  {journal} {\bibinfo  {journal} {Physical Review
  X}\ }\textbf {\bibinfo {volume} {11}},\ \bibinfo {pages} {031005} (\bibinfo
  {year} {2021})}\BibitemShut {NoStop}%
\bibitem [{\citenamefont {Samajdar}\ \emph {et~al.}(2021)\citenamefont
  {Samajdar}, \citenamefont {Ho}, \citenamefont {Pichler}, \citenamefont
  {Lukin},\ and\ \citenamefont {Sachdev}}]{samajdar2021quantum}%
  \BibitemOpen
  \bibfield  {author} {\bibinfo {author} {\bibfnamefont {R.}~\bibnamefont
  {Samajdar}}, \bibinfo {author} {\bibfnamefont {W.~W.}\ \bibnamefont {Ho}},
  \bibinfo {author} {\bibfnamefont {H.}~\bibnamefont {Pichler}}, \bibinfo
  {author} {\bibfnamefont {M.~D.}\ \bibnamefont {Lukin}}, \ and\ \bibinfo
  {author} {\bibfnamefont {S.}~\bibnamefont {Sachdev}},\ }\href@noop {}
  {\bibfield  {journal} {\bibinfo  {journal} {Proceedings of the National
  Academy of Sciences}\ }\textbf {\bibinfo {volume} {118}},\ \bibinfo {pages}
  {e2015785118} (\bibinfo {year} {2021})}\BibitemShut {NoStop}%
\bibitem [{\citenamefont {Samajdar}\ \emph {et~al.}(2020)\citenamefont
  {Samajdar}, \citenamefont {Ho}, \citenamefont {Pichler}, \citenamefont
  {Lukin},\ and\ \citenamefont {Sachdev}}]{samajdar2020complex}%
  \BibitemOpen
  \bibfield  {author} {\bibinfo {author} {\bibfnamefont {R.}~\bibnamefont
  {Samajdar}}, \bibinfo {author} {\bibfnamefont {W.~W.}\ \bibnamefont {Ho}},
  \bibinfo {author} {\bibfnamefont {H.}~\bibnamefont {Pichler}}, \bibinfo
  {author} {\bibfnamefont {M.~D.}\ \bibnamefont {Lukin}}, \ and\ \bibinfo
  {author} {\bibfnamefont {S.}~\bibnamefont {Sachdev}},\ }\href@noop {}
  {\bibfield  {journal} {\bibinfo  {journal} {Physical Review Letters}\
  }\textbf {\bibinfo {volume} {124}},\ \bibinfo {pages} {103601} (\bibinfo
  {year} {2020})}\BibitemShut {NoStop}%
\bibitem [{\citenamefont {Zhang}\ and\ \citenamefont
  {Cai}(2024)}]{zhang2024quantum}%
  \BibitemOpen
  \bibfield  {author} {\bibinfo {author} {\bibfnamefont {T.}~\bibnamefont
  {Zhang}}\ and\ \bibinfo {author} {\bibfnamefont {Z.}~\bibnamefont {Cai}},\
  }\href@noop {} {\bibfield  {journal} {\bibinfo  {journal} {Physical Review
  Letters}\ }\textbf {\bibinfo {volume} {132}},\ \bibinfo {pages} {206503}
  (\bibinfo {year} {2024})}\BibitemShut {NoStop}%
\bibitem [{\citenamefont {Hudomal}\ \emph {et~al.}(2022)\citenamefont
  {Hudomal}, \citenamefont {Desaules}, \citenamefont {Mukherjee}, \citenamefont
  {Su}, \citenamefont {Halimeh},\ and\ \citenamefont
  {Papi{\'c}}}]{hudomal2022driving}%
  \BibitemOpen
  \bibfield  {author} {\bibinfo {author} {\bibfnamefont {A.}~\bibnamefont
  {Hudomal}}, \bibinfo {author} {\bibfnamefont {J.-Y.}\ \bibnamefont
  {Desaules}}, \bibinfo {author} {\bibfnamefont {B.}~\bibnamefont {Mukherjee}},
  \bibinfo {author} {\bibfnamefont {G.-X.}\ \bibnamefont {Su}}, \bibinfo
  {author} {\bibfnamefont {J.~C.}\ \bibnamefont {Halimeh}}, \ and\ \bibinfo
  {author} {\bibfnamefont {Z.}~\bibnamefont {Papi{\'c}}},\ }\href@noop {}
  {\bibfield  {journal} {\bibinfo  {journal} {Physical Review B}\ }\textbf
  {\bibinfo {volume} {106}},\ \bibinfo {pages} {104302} (\bibinfo {year}
  {2022})}\BibitemShut {NoStop}%
\bibitem [{\citenamefont {Mukherjee}\ \emph
  {et~al.}(2020{\natexlab{c}})\citenamefont {Mukherjee}, \citenamefont {Sen},
  \citenamefont {Sen},\ and\ \citenamefont {Sengupta}}]{mukherjee2020dynamics}%
  \BibitemOpen
  \bibfield  {author} {\bibinfo {author} {\bibfnamefont {B.}~\bibnamefont
  {Mukherjee}}, \bibinfo {author} {\bibfnamefont {A.}~\bibnamefont {Sen}},
  \bibinfo {author} {\bibfnamefont {D.}~\bibnamefont {Sen}}, \ and\ \bibinfo
  {author} {\bibfnamefont {K.}~\bibnamefont {Sengupta}},\ }\href@noop {}
  {\bibfield  {journal} {\bibinfo  {journal} {Physical Review B}\ }\textbf
  {\bibinfo {volume} {102}},\ \bibinfo {pages} {075123} (\bibinfo {year}
  {2020}{\natexlab{c}})}\BibitemShut {NoStop}%
\bibitem [{\citenamefont {Verdeny}\ \emph {et~al.}(2016)\citenamefont
  {Verdeny}, \citenamefont {Puig},\ and\ \citenamefont
  {Mintert}}]{verdeny2016quasi}%
  \BibitemOpen
  \bibfield  {author} {\bibinfo {author} {\bibfnamefont {A.}~\bibnamefont
  {Verdeny}}, \bibinfo {author} {\bibfnamefont {J.}~\bibnamefont {Puig}}, \
  and\ \bibinfo {author} {\bibfnamefont {F.}~\bibnamefont {Mintert}},\
  }\href@noop {} {\bibfield  {journal} {\bibinfo  {journal} {Zeitschrift
  f{\"u}r Naturforschung A}\ }\textbf {\bibinfo {volume} {71}},\ \bibinfo
  {pages} {897} (\bibinfo {year} {2016})}\BibitemShut {NoStop}%
\bibitem [{\citenamefont {Weinberg}\ and\ \citenamefont
  {Bukov}(2017)}]{weinberg2017quspin}%
  \BibitemOpen
  \bibfield  {author} {\bibinfo {author} {\bibfnamefont {P.}~\bibnamefont
  {Weinberg}}\ and\ \bibinfo {author} {\bibfnamefont {M.}~\bibnamefont
  {Bukov}},\ }\href@noop {} {\bibfield  {journal} {\bibinfo  {journal} {SciPost
  Physics}\ }\textbf {\bibinfo {volume} {2}},\ \bibinfo {pages} {003} (\bibinfo
  {year} {2017})}\BibitemShut {NoStop}%
\bibitem [{\citenamefont {Weinberg}\ and\ \citenamefont
  {Bukov}(2019)}]{weinberg2019quspin}%
  \BibitemOpen
  \bibfield  {author} {\bibinfo {author} {\bibfnamefont {P.}~\bibnamefont
  {Weinberg}}\ and\ \bibinfo {author} {\bibfnamefont {M.}~\bibnamefont
  {Bukov}},\ }\href@noop {} {\bibfield  {journal} {\bibinfo  {journal} {SciPost
  Physics}\ }\textbf {\bibinfo {volume} {7}},\ \bibinfo {pages} {020} (\bibinfo
  {year} {2019})}\BibitemShut {NoStop}%
\end{thebibliography}%

\end{document}